\documentclass[sigconf]{acmart}

\AtBeginDocument{%
  \providecommand\BibTeX{{%
    \normalfont B\kern-0.5em{\scshape i\kern-0.25em b}\kern-0.8em\TeX}}}

\usepackage{enumitem}
\usepackage{subcaption}
\usepackage{caption}
\usepackage{xspace}
\usepackage{setspace}
\usepackage{tikz}

\newcommand{\ivtask}[1]{\textsc{task#1}}
\newcommand{\ivmethod}[1]{\textsc{method type#1}}
\newcommand{\ivcomp}[1]{\textsc{adjustment condition#1}}
\newcommand{\cours}[1]{\emph{Ours#1}}
\newcommand{\cauit}[1]{\emph{ParetoSelect#1}}
\newcommand{\cmanual}[1]{\emph{Manual#1}}

\newcommand{\anova}[6]{{\small [$F_{#1,#2}$\,$=$\,$#3$, $p$\,$#4$\,$#5$, $\eta_{g}^{2}$\,$=$\,$#6$]}} 
\newcommand{\pval}[2]{{\small ($p\,#1\,#2$)}} 
\newcommand{\pvall}[2]{{\small $p\,#1\,#2$}} 
\newcommand{\pvalr}[3]{{\small ($p\,#1\,#2, r = #3$)}} 

\copyrightyear{2025} 
\acmYear{2025} 
\setcopyright{cc}
\setcctype{by}
\acmConference[UIST '25]{The 38th Annual ACM Symposium on User Interface
Software and Technology}{September 28-October 1, 2025}{Busan, Republic of
Korea}
\acmBooktitle{The 38th Annual ACM Symposium on User Interface Software and
Technology (UIST '25), September 28-October 1, 2025, Busan, Republic of
Korea}\acmDOI{10.1145/3746059.3747645}
\acmISBN{979-8-4007-2037-6/2025/09}

\newcommand{\rev}[1]{\textcolor{black}{#1}}
\definecolor{blue}{rgb}     {0,0.0,1.0}

\begin{document}
\title{Preference-Guided Multi-Objective UI Adaptation}

\author{Yao Song, Christoph Gebhardt, Yi-Chi Liao, Christian Holz}





\affiliation{%
  \institution{Department of Computer Science}
  \institution{ETH Z\"urich, Switzerland}
  \city{} 
  \country{} 
}

\renewcommand{\shortauthors}{Song et al.}

\begin{abstract}

3D Mixed Reality interfaces have nearly unlimited space for layout placement, making automatic UI adaptation crucial for enhancing the user experience. 
Such adaptation is often formulated as a multi-objective optimization (MOO) problem, where multiple, potentially conflicting design objectives must be balanced.
However, selecting a final layout is challenging since MOO typically yields a set of trade-offs along a Pareto frontier. Prior approaches often required users to manually explore and evaluate these trade-offs, a time-consuming process that disrupts the fluidity of interaction.
To eliminate this manual and laborous step, we propose a novel optimization approach that efficiently determines user preferences from a minimal number of UI element adjustments. 
These determined rankings are translated into priority levels, which then drive our priority-based MOO algorithm. By focusing the search on user-preferred solutions, our method not only identifies UIs that are more aligned with user preferences, but also automatically selects the final design from the Pareto frontier; ultimately, it minimizes user effort while ensuring personalized layouts.
Our user study in a Mixed Reality setting demonstrates that our preference-guided approach significantly reduces manual adjustments compared to traditional methods, including fully manual design and exhaustive Pareto front searches, while maintaining high user satisfaction. We believe this work opens the door for more efficient MOO by seamlessly incorporating user preferences.

\end{abstract}



\begin{teaserfigure}
    \centering
    \includegraphics[width=\textwidth]{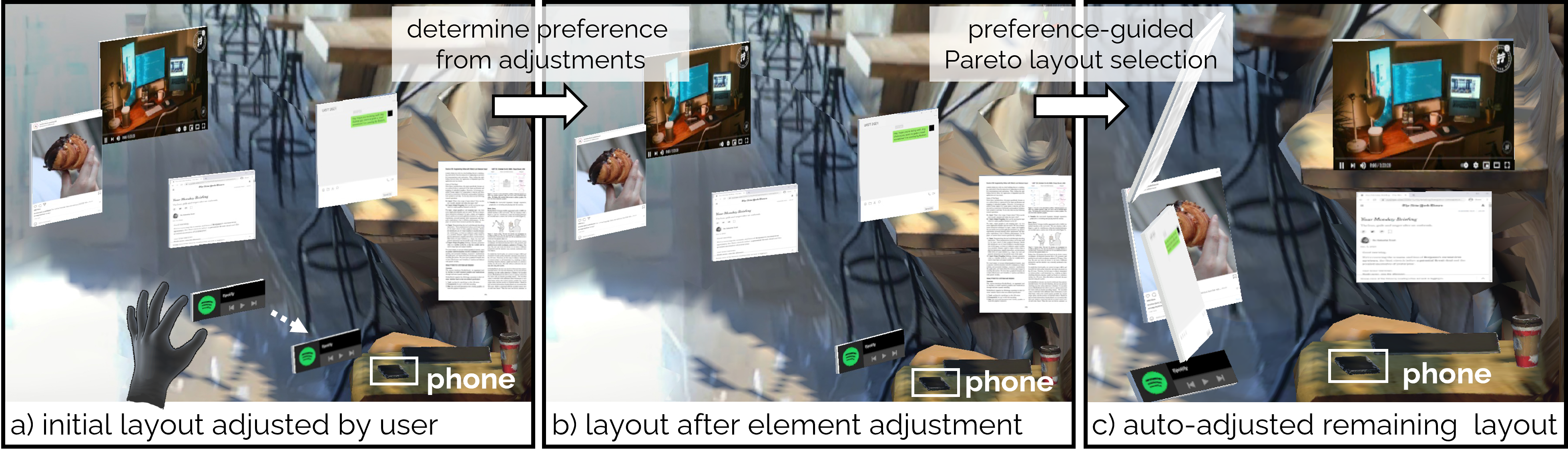}
    \caption{
    Our approach enables preference-guided multi-objective UI adaptation.
    In a coffee shop, an MR user wants to listen to music while watching videos.
    (a) After the user moves the virtual music player next to their phone, (a-b) our approach determines their preference for the objective of semantic agreement between virtual widgets (i.e., the music player) and physical objects (i.e., the phone) based on changes in objective values and prioritizes it over other terms.
    (b-c) It then defines a multi-objective optimization problem that conducts a Pareto search based on the set priority rank.
    Using the music player's new position as reference, it identifies the closest Pareto-optimal layout candidate by minimizing distance in objective space.
    (c) The selected layout adjusts all elements to maintain semantic agreement, positioning the video viewer and news page near the iPad, while aligning the messenger, Instagram, and music player closer to the phone.
    } 
    \Description{Sequence of a preference-based layout system: (a) User adjusts the initial layout of virtual windows in augmented reality, (b) System determines user preferences from these adjustments, (c) The system auto-adjusts the remaining elements to optimize layout based on user preferences using Pareto optimal selection.}
    \label{fig:teaser}
\end{teaserfigure}


\maketitle

\section{Introduction}
\label{sec:intro}


User Interfaces (UIs) are the fundamental touchpoint for interaction between users and computing systems, ranging from traditional 2D screens to advanced 3D environments in Mixed Reality (MR) scenarios. The adaptability of UIs in real-time to different user needs and preferences is crucial for enhancing a system's usability, usefulness, learnability, and user satisfaction, 
ultimately influencing its success and acceptance across various applications ~\cite{oulasvirta2020combinatorial}. 
To achieve adaptive UIs, previous work has formulated 2D UI layout assignment as a multi-objective optimization (MOO) problem to computationally identify promising layouts.
That is, using algorithms to explore various design candidates (i.e., different layouts) to find the ones that lead to optimal objective function values, considering multiple design objective functions (metrics to minimize or maximize).

For example, MOO in 2D GUIs can improve the order of menus by considering the frequency of usage and grouping~\cite{bailly2013menuoptimizer}, the sketching interface by optimizing for visual quality and selection efficiency~\cite{todi2016sketchplore}, or the layout of control panels governed by ergonomic criterion~\cite{uioptimizationgenetic2019diegomas}.
Researchers have also leveraged MOO to distribute UIs across multiple devices and users in collaborative environments~\cite{park2018adam}.
For UIs in 3D environments, adaptation is even more critical due to the vast space available for placing UI elements, where the quality of the layout can significantly impact user experience. Moreover, in an open 3D environment, manually adjusting every element is both costly and impractical.
In 3D UI layout design, common objective functions include semantic relationships~\cite{cheng2021semanticadapt}, ease of use of interaction methods~\cite{cheng2023interactionadapt}, muscle strain during interaction~\cite{johns2023towards}, field of view \cite{evangelista2022auit}, and more.
MOO enables the simultaneous consideration of these objectives, but they inherently involve trade-offs as improving one metric may negatively affect another. The result is typically a Pareto-optimal set of layouts, where each design excels in some objectives while compromising others.
While mathematically sound, this approach faces two major limitations that hinder its ability to fully adapt to diverse user needs and preferences.

The first challenge is \emph{the huge effort in manual evaluation on the Pareto-optimal designs}. 
As discussed, MOO produces a set of trade-off solutions (Pareto-optimal designs), but only one final outcome can be deployed in practice. Determining the final design remains an open question. Several approaches have been proposed.
One common approach is to transform a MOO problem into a single-objective problem by weighted sum \cite{cheng2021semanticadapt,cheng2023interactionadapt,evangelista2022auit}.
This solution leads to information loss since not all Pareto-optimal solutions can be reached when exploring solutions to a single objective.
Recent work allows users to \emph{exhaustively test} all the Pareto-optimal designs~\cite{johns2023towards}, and identify the most preferred outcome.
While this overcomes the limitations of the weighted sum approach, it significantly increases user effort, requiring them to engage with multiple UI layout candidates. This approach is time-consuming and disruptive to a streamlined user experience.
The second challenge is \emph{the neglect of user preferences in the MOO process}, which extends in two ways: First, Pareto frontier exploration is typically uniform across objective functions, without prioritizing user preferences. Given limited search time, this can lead to an explored frontier that fails to capture the \emph{true} Pareto frontier, potentially neglecting designs that better align with user needs. 
Second, when narrowing down to a final design, the process is often disconnected from a concrete estimate of user preferences, leading to subjectively suboptimal outcomes.

To overcome these limitations, we propose a novel optimization approach for UI layout adaptation that \emph{searches the Pareto frontier guided by the user's preference} while \emph{identifying the final design candidate based on the determined user preference}.
Our approach, combines three features:
(a)~The system determines the users' preferences for design objectives in a human-in-the-loop manner, directly from their interaction with the UI,
(b)~our method uses these preferences to guide the dynamic Pareto frontier exploration
and (c)~automatically derives the most preferred design.

\rev{Figure~\ref{fig:teaser} illustrates our approach. Each iteration begins with the user adjusting a UI element in an initial layout. From this input, we determine the user's preference ranking over design objectives by analyzing how the adjustment impacts objective function values.
Based on this analysis, objectives are grouped into priority tiers. 
These rankings are then used by our mixed-Pareto-lexicographic MOO approach \cite{lai2021solving} to guide the Pareto-front search and identify optimal designs within a fixed search budget. 
This enables more personalized and preference-aligned results than traditional MOO approaches, which assume fixed Pareto-optimal sets.
Finally, the system selects the layout closest to the user’s adjustment in objective space.
Our framework, illustrated in \autoref{fig:flowchart}, iterates through this process until the user is satisfied with the layout.
It introduces two key distinctions from prior work: (1) it eliminates the need for users to manually explore all Pareto-optimal designs by identifying a candidate based on few user-adjusted elements; (2) it leverages inferred user preferences to guide the Pareto search, enabling personalized, preference-driven UI adaptation.}

We demonstrate our approach within a 3D MR context, but our method can equally be applied to 2D UIs. 
To evaluate its effectiveness and efficiency, we conducted a user study to compare it against two established baselines: manually adjusting the elements of an initial Pareto-optimal layout, and selecting the Pareto-optimal layout closest to the manually adjusted UI element positions.
%
The results showed that our approach significantly reduced the need for manual UI adjustments while maintaining high user satisfaction with the final design.
While overall user experience ratings did not show a significant difference from baseline methods, the results indicate promising trends.
Notably, the p-values for both metrics were relatively low, with ours consistently trending better than baselines across criteria.
These results can be attributed to our approach’s ability to conduct a more focused search, as evident by a smaller objective space distance to user-specified references for prioritized objectives.
At the same time, the equivalent hypervolume compared to the baseline indicates that this increased focus does not come at the cost of overall design space exploration.


In summary, our contributions are as follows:

\begin{itemize}[noitemsep,leftmargin=*]
    \item \rev{A novel inverse approach that derives the user's preference ranking of objective functions from their UI layout manipulations.}
    
    \item \rev{The first application of mixed Pareto-lexicographic MOO for UI adaptation, leveraging ranked design objectives to model user preferences, effectively guiding a Pareto-frontier search to identify the most preferred UI layouts.}
    
    \item An empirical study ($N=12$) demonstrating that, compared to two baseline methods, our approach produced layouts that better matched participants' preferences while requiring less effort to personalize.
\end{itemize}

\section{Background \& Related work}
\label{sec:related_work}

Our proposed approach builds on two key concepts: Pareto-optimal solutions in MOO problems and the Lexicographic MOO method. 
In this section, we provide an overview of the MOO problem, which leads us to Pareto-optimal solutions, and motivates the need for Lexicographic MOO. 

\subsection{Multi-Objective Optimization (MOO)}

\subsubsection{MOO problem}

MOO involves simultaneously optimizing multiple objective functions, aiming to identify optimal combinations of decisions across objectives.
The general mathematical form of an MOO problem is given by: 
\begin{equation*}
\begin{aligned}
& \text{min} \quad \{f_1(\mathbf{x}), f_2(\mathbf{x}), \ldots, f_k(\mathbf{x})\} \\
& \text{subject to} \quad g_j(\mathbf{x}) \leq 0, \quad j = 1, 2, \ldots, m, \\
& \quad \quad \quad \quad \quad h_l(\mathbf{x}) = 0, \quad l = 1, 2, \ldots, e \\
& \quad \quad \quad \quad \quad \mathbf{x} \in X \\
\end{aligned}
\end{equation*}
where $k$ represents the number of objective functions, $m$ and $e$ represent the number of inequality and equality constraints respectively. The variable $\mathbf{x}$ represents the decision variable within the design space $X$.

The challenge of a MOO problem is that different objectives may not be fully aligned with each other, and that they could even be in conflict.
That is, a design decision, $x$, is good for an objective (e.g., $f_1$) but may be a poor decision for another design objective (e.g., $f_2$).
Due to such a multi-objective nature, an MOO problem usually can not conclude with one optimal solution. Instead, it yields a set of solutions where improving one objective often comes at the cost of worsening another~\cite{marler2004survey}.

These so-called Pareto-optimal solutions are the best trade-offs in this objective space. 
For more details, we refer the reader to \citeauthor{gunantara2018review}'s introduction to Pareto optimality and MOO~\cite{gunantara2018review}.


In the context of UI design and adaptation, the MOO problem is framed within a design space characterized by design factors and a set of objectives that map combinations of these design factors to an objective space \cite{hofmann2023optimism}. 
This enables the identification of optimal design factors given the objectives.
A common use case is optimizing UI layouts in Virtual Reality.
In this context, an optimizer can, for example, determine the positions and dimensions of UI elements, balancing the trade-off between visibility and reachability, as seen in tools like AUIT\cite{evangelista2022auit}.

\subsubsection{Solvers for MOO problems}

Several types of optimization methods have been proposed to explore and identify the Pareto frontier in the objective space. 
Evolutionary algorithms are commonly used for the search process because they are gradient-free, allowing them to handle differentiable and non-differentiable objective functions.
Within evolutionary algorithms, the Non-dominated Sorting Genetic Algorithm II (NSGA-II) ~\cite{deb2002fast} is a common selection to explore the Pareto frontier for its computational efficiency. 
Below, we briefly provide the working principles of the NSGA-II algorithm:
\begin{enumerate}[leftmargin=*]
    \item \textbf{Initialization and Objective Evaluation}: A population of candidate solutions is generated through random sampling, and each solution is evaluated based on multiple objectives.
    
    \item \textbf{Non-dominated Sorting}: Solutions are ranked into non dominated fronts based on Pareto dominance. Solutions in the first front dominate all others, while subsequent fronts contain increasingly dominated solutions.
    
    \item \textbf{Crowding Distance and Selection}: Crowding distance is used to maintain diversity within each front. For non-boundary solutions, it is calculated as the normalized difference between the objective values of adjacent solutions. Boundary solutions (those with the smallest and largest values for each objective) are assigned an infinite crowding distance to ensure their preservation ~\cite{deb2002fast}. Solutions are selected based on front ranking (prioritizing lower-ranked fronts) and crowding distance (favoring more diverse solutions).
    
    
    \item \textbf{Genetic Operations (i.e., Crossover and Mutation)}: Selected solutions perform genetic operations (usually, crossover and mutation) to create offspring that explore new areas of the solution space.
    
    \item \textbf{Elitism and Population Update}: The parent and offspring populations are combined, and the combined set is sorted again into non-dominated fronts. The best solutions are retained for the next generation, ensuring that the elite solutions are carried forward. The process repeats until a stopping criterion is met.
\end{enumerate}


\subsection{Mixed Pareto-Lexicographic Multi-Objective Optimization (MPL-MOO)}
\subsubsection{MPL-MOO problem}
The standard MOO problem assumes equal importance for all objectives, leading to evenly distributed Pareto-frontier searches across the objective space. However, in practice, this may not reflect user preferences.
To tackle this limitation, the class of Mixed Pareto-Lexicographic Multi-Objective Optimization problems has been introduced. 
The generic formulation of the MPL-MOO problem is

\begin{equation*}
\begin{aligned}
& \text{min} \quad \{f_1(\mathbf{x}), f_2(\mathbf{x}), \dots, f_k(\mathbf{x})\} \\
& \text{subject to} \quad \wp(f_1, f_2, \dots, f_k)
\end{aligned}
\end{equation*}

where $\wp(\cdot)$ is a general priority distribution among the objectives.
In practice, this priority distribution is either realized as priority chains (PC-MPL-MOPs) \cite{lai2020solving} or as priority levels (PL-MPL-MOPs) \cite{lai2021solving}.
In this paper, we build on the latter. 
The general mathematical structure of the PL-MPL-MOP is

\begin{equation*}
\label{lex}
\begin{aligned}
\text{lexmin} \left[ \min \left( \begin{array}{c}
f_1^{(1)}(\mathbf{x}) \\
f_2^{(1)}(\mathbf{x}) \\
\vdots \\
f_{k_1}^{(1)}(\mathbf{x})
\end{array} \right) , 
\min \left( \begin{array}{c}
f_1^{(2)}(\mathbf{x}) \\
f_2^{(2)}(\mathbf{x}) \\
\vdots \\
f_{k_2}^{(2)}(\mathbf{x})
\end{array} \right) , 
\dots , 
\min \left( \begin{array}{c}
f_1^{(p)}(\mathbf{x}) \\
f_2^{(p)}(\mathbf{x}) \\
\vdots \\
f_{k_p}^{(p)}(\mathbf{x})
\end{array} \right) 
\right]
\end{aligned}
\end{equation*}

where $p$ is the number of priority levels and $f_i^{(j)}$ is the $i$th objective in the $j$th level of importance. 
The $lexmin$ operator indicates that the minimization of the objectives is ranked lexicographically, meaning that objectives $f_i^{(1)}$ are infinitely more important than $f_i^{(2)}$, which in turn have priority over $f_i^{(3)}$, and so on.

The assumption of PL-MPL-MOO is that some groups of objectives have clear precedence over some others in a lexicographic sense.
In our work, we leverage this formulation by forming priority groups for objectives based on user preferences determined from their adjustments to an existing MR UI layout.
This enables Pareto-frontier searches across the objective space that are guided by users' preferences leading to designs that are better tailored to individual needs.

\subsubsection{Solvers for PL-MPL-MO problems}

To solve PL-MPL-MO problems, several adaptations of NSGA-II have been proposed for priority-guided Pareto frontier exploration, such as PL-NSGA-II and Lex-NSGA-II \cite{lai2021solving}.
In our work, we employ PL-NSGA-II, which integrates priority levels by performing Pareto dominance checks within each priority level and conducting lexicographic domination checks across priority levels. Here we provide the high-level working principles of PL-NSGA-II:

\begin{enumerate}[leftmargin=*]
    \item \textbf{Initialization and Objective Evaluation}: A population of candidates solutions is generated, and each solution is evaluated based on multiple objectives, as in NSGA-II.
    
    \item \textbf{Non-dominated Sorting}: Solutions are divided into Pareto non-dominated subfronts within each priority level, ensuring a hierarchical ranking based on their priority structure. Within each priority level, the first subfront consists of solutions that dominate all others, while subsequent subfronts contain solutions that are progressively more dominated.
    
    \item \textbf{Lexicographic Ordering and Crowding Distance}: 
    Solutions are first compared based on higher-priority objectives. If they are equivalent in higher-priority objectives, the algorithm proceeds to evaluate lower-priority objectives using lexicographic ordering (see \autoref{lex}). To maintain diversity, the crowding distance is computed within each non-dominated subfront at a given priority level.
    
    
    \item \textbf{Selection and Genetic Operations}: Solutions are selected for reproduction based on their non-dominated rank, lexicographic preferences, and crowding distance. These selected solutions perform crossover and mutation to generate new offspring.
    
    \item \textbf{Elitism and Population Update}: The parent and offspring populations are combined, and solutions are re-ranked using non-dominated sorting and lexicographic ordering. The best solutions are retained for the next generation. The process repeats until a termination condition is satisfied.
\end{enumerate}

To summarize, key differences are two-fold:
(i) \textbf{Lexicographic Ordering}: After non-dominated sorting (i.e., identifying Pareto-optimal solutions) within each priority level, PL-NSGA-II employs a lexicographic hierarchy across priority levels, where higher-priority objectives always take precedence over lower-priority ones.
(ii) \textbf{Selection Criteria}: PL-NSGA selects solutions not only based on Pareto dominance and crowding distance but also by considering user-defined objective preferences, making it better for cases with objective hierarchies.

\begin{figure*}[t]
  \centering
  \includegraphics[width=1.\linewidth]{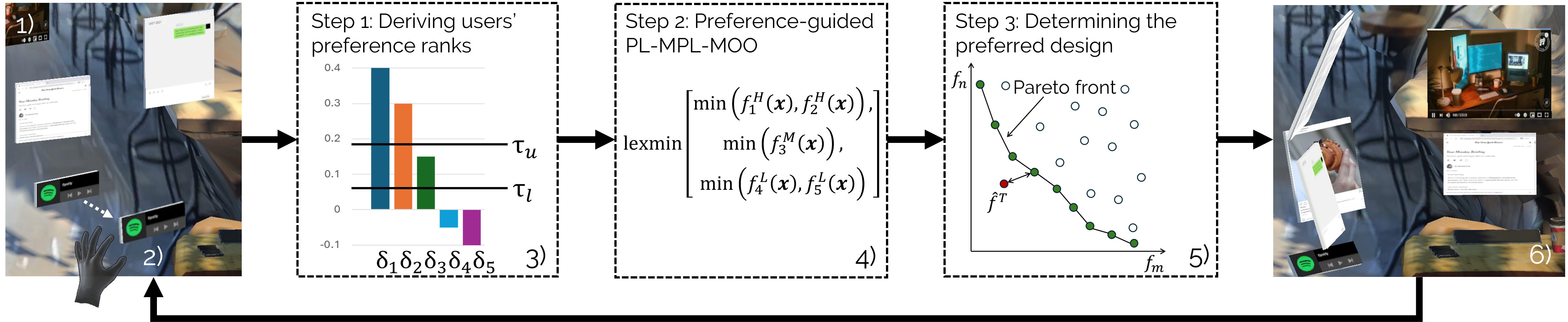}
  \caption{Overall pipeline of our approach: 1) After presenting an initial UI layout, 2) the user can move UI elements according to their preference. 3) Based on these movements, we deduce their preference for optimization objectives based on differences of layouts in the objective space ($\delta_1, ..., \delta_5$, Section \ref{sec:deduduce-preference}). 4) Via thresholding ($\tau_u, \tau_l$), we then identify priority levels among objectives based on which we formulate a PL-MPL-MOP and perform a preference-guided dynamic Pareto frontier exploration (Section \ref{sec:preference-guided-pareto-search}). 5) Finally, we determine the preferred Pareto-optimal design by using a user's accumulated UI element movements as a reference point $\hat{f}^{T}$ and selecting the design on the Pareto frontier with minimal distance (Section \ref{sec:determining-optimal-design}). (6) The adjusted layout is presented to the user, who can then repeat the process until satisfied with the final design.}
  \Description{
  The figure illustrates our approach for personalized UI layout adaptation in three steps.
    The user interacts with the MR environment, adjusting UI elements.
    The system captures these adjustments to determine user preferences.
    Step 1: The user's preferences are derived by analyzing how their modifications influence objective function values, assigning preference ranks.
    Step 2: Using the determined preference ranks, a preference-guided Multi-Objective Optimization (MOO) problem is formulated via lexicographic ordering to guide the search for Pareto-optimal solutions.
    Step 3: The system determines the preferred design by selecting the closest candidate from the Pareto frontier that aligns with the user's preferences.
    The final UI layout reflects the user's preferred configuration, with adjusted element positions based on their input and optimization results.
  }
  \label{fig:flowchart}
\end{figure*}

\subsection{Preference-guided MOO in Human-Computer Interaction}
MOO problems typically yield a set of Pareto-optimal solutions, each representing different trade-offs among the objectives. Despite the availability of multiple Pareto-optimal designs, only one solution is usually selected in practice. Selecting a final design from the entire Pareto frontier remains an open research challenge since no single Pareto-optimal design is mathematically superior to the others. Previous research in Human-Computer Interaction has attempted to address this challenge by offloading it onto the user. 

One intuitive approach to resolving the challenge of multiple solutions is to transform the MOO problem into a single-objective problem using weighted sums. This method converts the problem into identifying the most suitable weight setting among the design objectives. Numerous studies have explored allowing users and designers to manually adjust these weights to determine the ideal configuration through trial and error. For instance, MenuOptimizer \cite{bailly2013menuoptimizer} provides an Objective Space Panel where designers can set weights for optimization objectives such as performance, consistency, and similarity. Similarly, OPTIMISM \cite{hofmann2023optimism} introduces a GUI to help designers and end-users specify objective and heuristic weights for optimization in fabrication tasks. In online 3D UI layout adaptation, tools like SemanticAdapt \cite{cheng2021semanticadapt}, InteractionAdapt \cite{cheng2023interactionadapt}, SituationAdapt \cite{li2024situationadapt}, and AUIT \cite{evangelista2022auit} offer control panels within Unity for adjusting weights linked to various objectives. Recently, \citet{10.1145/3613904.3642071} explored eliciting population-level optimal objective weights based on users' subjective ratings of Pareto-optimal designs. While this method is simple and effective, reducing multiple objectives into a single one inevitably leads to information loss, as certain designs on the Pareto frontier may not be discovered in the weighted-sum solution.

To mitigate this limitation, other approaches have been proposed to explore the Pareto frontier and conclude a final solution based on user preferences. For example, ParetoAdapt \cite{johns2023towards} presents users with the visualization of the Pareto frontier, allowing them to freely explore and select their preferred Pareto-optimal designs. Instead of focusing on a single Pareto-optimal point, prior research has also explored clustering designs based on similar objective values. For instance, \citet{chan2022investigating} employed multi-objective Bayesian optimization for a 3D input interaction task to find the best trade-offs between accuracy and efficiency. By clustering designs based on objective values, users could choose a design that aligns with their preference for higher accuracy or efficiency. Similar works include clustering haptic device designs based on information transfer and recognition accuracy \cite{liao2023interaction}, or clustering other input techniques based on various performance metrics \cite{chan2022investigating}, offering users the option to prioritize specific objectives.

While these methods have advanced the selection of a final design in MOO problems based on user preferences, they have three major limitations. First, users must select directly based on objective values, which are often difficult for humans to interpret intuitively. Second, the process involves extensive trial and error, making it time-consuming, tedious, and prone to biases and noise. Finally, the derivation of the Pareto frontier typically occurs independently of user preferences, even though user preferences could provide crucial insights for guiding the Pareto frontier search in more relevant directions.
Our approach addresses these limitations by eliminating the need for users to manually select designs based on objective values. Instead, we identify the most preferred design based on users' manipulation of UI elements. This method not only alleviates the challenge of interpreting objective values but also removes the need for repeated trial and error. Moreover, by leveraging the PL-MPL-MOO algorithm, our workflow enables a Pareto frontier search guided by the priority of objectives determined from users' UI interactions, leading to more accurate and customized Pareto frontier exploration.


\section{Approach}
\label{sec:Method}
Our approach identifies user preferences for objectives and leverages them to conduct a guided exploration of the Pareto frontier, enabling the identification of designs that more closely align with user preferences.
Figure~\ref{fig:flowchart} illustrates the approach, which consists of three steps: 
(a) deducing users' preferences for design objectives,
(b) preference-guided dynamic exploration of the Pareto frontier,
(c) and automatically identifying the most preferred Pareto-optimal design.
Below, we provide a detailed explanation of each step.

\subsection{Step~1: Deriving users' preference ranks for design objectives}
\label{sec:deduduce-preference}
In the first step, we determine user preference ranks over objective terms from their UI element adjustments to an initially generated layout.
Specifically, by computing the differences in objective function values before and after user adjustments, we deduce their preferences for design objectives based on the subset of moved elements among the total \( N \)  UI elements.

Our core assumption is that a user's preferences for specific design objectives are reflected in the amount of changes made to those objective values. 
For example, if a user relocates an element in a way that significantly reduces neck strain, we determine that minimizing neck strain has a higher priority than expressed in the current layout, and the optimization process should prioritize this objective. 
Conversely, if a user moves an element to a position that worsens its reachability, it suggests that reachability has a reduced importance, and should therefore receive less emphasis in the optimization.
Building on this, we further assume that the UI layout after user adjustments reflects their preferred state at that moment. With each iteration, user modifications progressively refine the layout, guiding the optimization process toward a final design that best aligns with the user's implicit preferences.

Note that a full UI adaptation may take multiple iterations, with each iteration representing a complete loop in \autoref{fig:flowchart}.
A straightforward solution is seeing each iteration as an independent event; yet, focusing solely on the changes in the current UI movements may introduce noise and miss the information gained from previous user adjustments. 
To address this, we adopt a more robust approach by storing all changes to objective values and applying weighted importance to different iterations—recent iterations receive higher weights, while earlier ones receive lower weights. 
This method enables us to fully utilize prior information while smoothing out noisy human adjustments.


Formally, we define the initial UI positions as \([x_i, ..., x_M]\) with the corresponding objective values \(f_j = f([x_i, ..., x_M])\).
The UI positions of the elements the user has moved are defined as \([x_i^{\hat{}}, ..., x_M^{\hat{}}]\) with the corresponding objective values \(f_j^{\hat{}} = f_j([x_i^{\hat{}}, ..., x_M^{\hat{}}])\).
Here, $M$ denotes the number of moved UI elements.
To get a better estimate of users' preferences, we consider multiple layout adjustments, where $f_j^t$ and $\hat{f_j}^t$ are computed for each adjustment $t \in \{1, 2, \dots, T\}$, with $T$ representing the final layout adjustment.
Using these objective values, we then compute the difference between objective values $\delta_{j}^t$ as 

\begin{equation}
\delta_{j}^t = \frac{\hat{f_j}^t - f_j^t}{f_{j}^{\max}}, \quad \text{where } j \in \{1, 2, \dots, K\}. 
\end{equation}

\noindent Here, $f_{j}^{\max}$ represents the maximal value the objective function $f_j$ can return and $K$ denotes the total number of objectives. 
This normalization is a common practice to maintain a consistent objective range, and similar approaches are used in prior human-in-the-loop optimization studies \cite{chan2022investigating,johns2023towards}.
We then compute the Triangular Moving Average to obtain $\delta_{j}$, representing the difference in values for each objective function in the history of layout adjustments:
\begin{equation}
\delta_j = \frac{\sum_{t=1}^{T} t \cdot \delta_{j}^t}{\sum_{t=1}^{T} t}~.
\end{equation}

The variable $\delta_j$ reflects whether the user's UI element movements indicate an increase or decrease in preference for the objective function $f_j$. 
By weighting $\delta_{j}^t$ based on its position in the adjustment history, we prioritize recent adjustments closer to the final adjustment $T$, while gradually reducing the influence of earlier adjustments.

\subsection{Step~2: Preference-guided dynamic Pareto frontier exploration}
\label{sec:preference-guided-pareto-search}

In our second step, we incorporate user preferences into the MOO problem by formulating it as PL-MPL-MOP to enable a preference-guided Pareto search.
To this end, we use $\delta_j$ to determine the PL-MPL-MOP's priority levels. Specifically, we define the priority groups as follows:

\begin{equation}
\begin{aligned}
H &= \{ j \mid \delta_j > \tau_u \} \\
M &= \{ j \mid \tau_l \leq \delta_j \leq \tau_u \} \\
L &= \{ j \mid \delta_j < \tau_l \}
\end{aligned}
\end{equation}

where \(\tau_l\) and \(\tau_u\) are thresholds for the lower and upper priority level respectively.
The final objective function of our problem is then given by

\begin{equation}
\begin{aligned}
\text{lexmin} \left[ \min \left( \begin{array}{c}
f_1^{(H)}(\mathbf{x}) \\
f_2^{(H)}(\mathbf{x}) \\
\vdots \\
f_{k_1}^{(H)}(\mathbf{x})
\end{array} \right) , 
\min \left( \begin{array}{c}
f_1^{(M)}(\mathbf{x}) \\
f_2^{(M)}(\mathbf{x}) \\
\vdots \\
f_{k_2}^{(M)}(\mathbf{x})
\end{array} \right) , 
\min \left( \begin{array}{c}
f_1^{(L)}(\mathbf{x}) \\
f_2^{(L)}(\mathbf{x}) \\
\vdots \\
f_{k_p}^{(L)}(\mathbf{x})
\end{array} \right) 
\right]
\end{aligned}
\end{equation}

where $H$, $M$, and $L$ define the group of objective functions with high, mid, and low priority levels, respectively.
The decision to establish three groups allows us to classify objectives into prioritized (important or preferred), neutral (no strong preference), and deprioritized (not preferred) categories. This tri-level classification provides a simple yet robust way to capture user preferences within the objective space.
To solve this problem, we use PL-NSGA-II to perform Pareto frontier searches within the objective space, guided by the defined priority groups.
This ensures that objectives in $H$ take precedence over those in $M$, and that those in $M$ take precedence over $L$.
Since the priority groups are derived from users' UI element adjustments, this approach helps identify Pareto optimal designs that are better aligned with user preferences.

\subsection{Step~3: Identifying the most preferred Pareto-optimal layout}
\label{sec:determining-optimal-design}

In the final step of our approach, we leverage the user-adjusted UI elements (element \(  i \) to element \(  M \)) to identify the preferred design within the Pareto frontier. 
Specifically, we map the new positions of the adjusted UI elements \([\hat{x_i}^T, ..., \hat{x_M}^T]\) to the objective space to attain a user-defined reference point  \(\hat{f^T} = [f_1([\hat{x_i}^T, ..., \hat{x_M}^T]), ...,\) \(f_K([\hat{x_i}^T, ..., \hat{x_M}^T])]\). 
We then search through the set of Pareto optimal layouts $X$ to find the point \( f^* \) with minimal distance: 

\begin{equation}
\min_{[x_i, ..., x_M] \subset X} d\left( f, \hat{f^T} \right), 
\label{eq:distance-minimization}
\end{equation}

\noindent where $d$ is a function that computes the two-norm distance between the reference point $\hat{f^T}$ and a Pareto-optimal design $f = [f_1([x_i, ..., x_M]), ..., f_K([x_i, ..., x_M])]$, only considering the positions of the $M$ explicitly adjusted UI elements. 
Our approach selects the minimal-distance solution $[x_i^{*}, ..., x_M^{*}]$ and updates the entire layout $[x_1^{*}, ..., x_N^{*}]$ accordingly.

\section{Example use case: Mixed Reality UI layout adaptation}
\label{sec:Implementation}
While our approach is applicable to both traditional 2D user interfaces and 3D MR applications, \rev{we focus on an MR use case, a common application of adaptive user interfaces \cite{cheng2021semanticadapt,cheng2023interactionadapt,evangelista2022auit}, to demonstrate its capabilities.} 
Thus, we implemented an MR environment (\autoref{fig:teaser}) with a multi-widget UI, similar to those provided by Apple Vision Pro or Meta Quest. 

\subsection{Objective Terms}
\label{sec:objective-terms}
In the use-case, we aim to utilize our approach to adapt the layout of an MR UI featuring seven virtual widgets. 
To optimize the layout, we focus on five \rev{well-established} objectives drawn from previous research:
minimizing neck strain \cite{johns2023towards}, minimizing shoulder load \cite{johns2023towards}, balancing visibility and reachability \cite{cheng2021semanticadapt}, ensuring placement in the field of view and enhancing semantic agreement with environmental objects \cite{cheng2021semanticadapt}.
All objective functions are defined as cost functions based on the UI elements' positions.
\rev{Note that our framework is extensible to other virtual widgets and objectives.}
In the following, we explain the impact of the five objective functions on an MR layout (refer to the respective paper for their equations):

\begin{description}
    \item \textbf{Minimizing neck strain} \cite{johns2023towards}: 
   For each virtual widget, determine the angular distance between the user's eye and the widget’s position in the x-z plane.
   Multiply this angular distance by the widget's observation probability. A higher cost signifies that widgets with greater observation probability are positioned above or below the user's eye level, while a lower cost indicates alignment closer to eye level.
    
    \item \textbf{Minimizing shoulder load} \cite{johns2023towards}: 
    Similar to the neck strain objective, calculate the angular distance between the user's shoulder and each virtual widget's position.
    Multiply this distance by the widget's interaction probability. A higher cost signifies that frequently interacted widgets are positioned substantially above or below the user's shoulder level, while a lower cost indicates placement closer to the shoulder, facilitating easier interaction.
    
    \item \textbf{Balancing visibility and reachability} \cite{cheng2021semanticadapt}: 
    Based on viewing ergonomics \cite{ankrum1995posture}, this objective is minimized when the UI element is positioned 0.5 meters from the user and maximized at 0 or infinite distance.
    An exponential interpolation function determines the cost for intermediate distances.
    
    \item \textbf{Ensuring field of view placement} \cite{cheng2021semanticadapt}: 
    For each virtual widget, compute the angular difference between its position and the user's gaze direction, then weight it by the widget's observation probability.
    If this angular difference is less than 5 degrees (i.e., within the foveal region), the cost is zero.
    A larger angular difference incurs a higher cost (off to the side or behind the user), signifying the widget is farther from the user's gaze, whereas a smaller difference results in a lower cost, indicating proximity to the foveal area (directly in front of the user).
    
    \item \textbf{Semantic agreement} \cite{cheng2021semanticadapt}:
    For each virtual widget, compute its distance to all defined real-world objects around the user and weight it by a semantic association cost derived from WordNet \cite{miller1995wordnet}.
    This approach aligns UI elements with real-world objects based on semantic meaning.
    A high semantic association cost and/or a large distance lead to higher overall costs, while weak semantic associations or short distances result in lower costs.
\end{description}

\subsection{Use-case Scenario}
To better illustrate how our approach can support users in a multi-widget MR environment based on the defined objectives, we describe the following scenario:
A user is sitting in a coffee shop, wearing an MR headset, and opening various applications on the device (as shown in \autoref{fig:teaser}). 
Physical objects such as a smartphone, an iPad, and a cup of coffee are placed on the table.
The visible virtual UI elements include Instagram, a video viewer, a messenger, a paper reader, one news web page, and a music player.
Having finished reading the paper, the user now wants to switch to other activities, such as listening to music or watching a video. The user prefers to have entertainment-related applications closer and the paper to the side of the view area. To articulate this preference, the user moves the virtual music player next to the phone (\autoref{fig:teaser}a). 

Based on the user's adjusted layout (Figure \ref{fig:teaser}b), our approach determines their preference for semantic agreement and adjusts the entire layout accordingly (Figure \ref{fig:teaser}c).
The music player is positioned next to the phone, the video viewer is centered, and the paper is placed to the side. Additionally, the video viewer and news pages are moved closer to the iPad, while the messenger, Instagram, and music player are aligned nearer to the phone to achieve higher semantic agreement.
The layout also minimizes neck strain while ensuring good reachability and visibility of the UI widgets.

\subsection{Implementation}
We implemented this MR showcase in Unity by building on the implementation of SemanticAdapt  \cite{cheng2021semanticadapt}.
MR environments are scanned real-world 3D rooms where each UI element is depicted as a virtual widget in 3D space (see Figure \ref{fig:enviroments}). 

Our adaptation approach is implemented in Python 3.9 using the Pymoo package \cite{blank2020pymoo}.
Specifically, we adapted their NSGA-II implementation to develop a PL-NSGA-II solver.
We computed the Pareto frontier with a population size of 100 across 80 generations. We set the thresholds for the priority levels to \(\tau_l  = 0.0\) and \(\tau_u  = 0.2\), which were determined via a 3-participant pilot test.

The Unity application communicated with our Python optimizer via websockets.

\section{Evaluation}
\label{sec:evaluation}

To evaluate if our approach generates MR layouts that better align with users' preferences, we compare our approach with two baseline adaptation methods.
Our study aims to address the following two research questions: 
(1) Can our approach generate layouts that more accurately align with users' preferences? 
(2) Consequently, can it reduce the number of element movements users need to personalize a layout?

\subsection{Study Design}
We used a within-subject design with two variables: \ivtask{} (2 levels: free personalization, targeted personalization) and \ivmethod{} (3 levels: \cauit{}, \cmanual{}, \cours{}).
We also varied the context in which a task was performed by changing the environment (coffee shop, home office).
The order of \ivtask{}s was fixed, with free personalization always presented first and targeted personalization second. We applied Latin-square counterbalancing to determine the presentation order of \ivmethod{} and to assign the environment for each \ivtask{}.
The independent variables are explained in the paragraphs below.

In each condition, we collected usability- and user-experience-related metrics such as the number of elements participants explicitly moved, and participants' perceived task load using the NASA-TLX \cite{Hart1986}. 
After each \ivtask{}, participants were further asked to rank the layouts generated by each \ivmethod{} after making their adjustments. 
Lastly, we asked for participants' overall preference for \ivmethod{} and had them complete the UEQ-S \cite{schrepp2023comparison}.

\rev{To compare between MOO-based approaches (i.e., \cauit{} and \cours{})}, we measured the average distance between values of participants' moved UI elements in objective space and the nearest solution candidate on the Pareto frontier, using $d(f, \hat{f^T})$ as defined in Equation \ref{eq:distance-minimization}. 
Additionally, we recorded the average objective values.
For both metrics, we differentiated between objectives in the highest priority group ($H$) and across all priority groups ($H$, $M$, $L$). 
Lastly, we calculated the hypervolume across all priority groups. 
Hypervolume is a widely used metric for assessing the results of a Pareto-frontier exploration or MOO. It measures the volume in a multi-dimensional space dominated by the Pareto-optimal points, relative to a given reference point. 
A higher hypervolume indicates that better overall solutions are identified. For details about the definition and calculation of hypervolume, we refer to Section 2 of \citet{shah2016pareto}.

\paragraph{Tasks} The study consisted of two tasks: 1) free personalization and 2) targeted personalization, presented to participants in that order.
In free personalization, participants were allowed to freely customize the UI layout according to their preference.
In the targeted personalization task, participants were first introduced to the objective functions outlined in Section \ref{sec:objective-terms} (the explanation of terms used in the experiment is detailed in Appendix \ref{app:task-explanations}).
They were then asked to prioritize the defined terms and make adjustments to the elements, so that movements would align with the objectives of the terms that were more important to them.
Both tasks ended when the participants were satisfied with their customized layout.

As our approach models preference in the objective space, it can only estimate preferences related to the objective terms defined in the optimization problem. 
To evaluate its performance in a controlled setting, we designed the second task to constrain user preferences to these predefined objectives.
In contrast, the first task simulates a more realistic scenario where not all user preferences are explicitly modeled, allowing us to assess whether our approach remains effective in such cases.
\rev{We fixed the task order to always begin with free personalization to avoid priming participants with the predefined objectives of the targeted personalization task.}

\begin{figure}[t]
  \centering
  \includegraphics[width=1.0\linewidth]{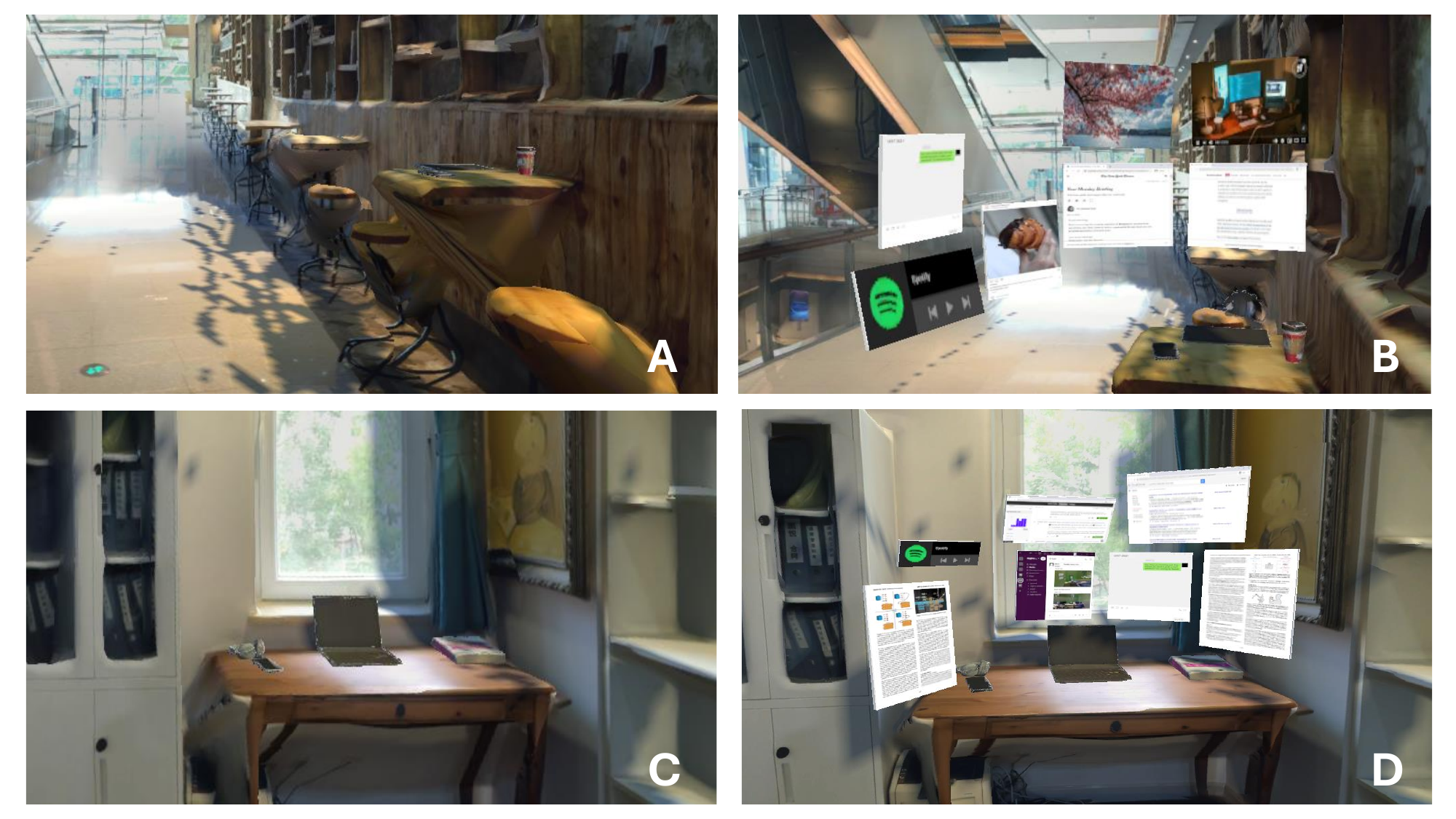}
  \caption{Study environments: (A) coffee shop, (B) coffee shop with initial layout, (C) home office, (D) home office with initial layout.}
  \Description{Comparison of virtual workspaces: (A) A physical café environment, (B) A café with augmented digital workspace including floating windows and media, (C) A minimalist home office, and (D) The same home office enhanced with augmented reality displays showing various applications and content in floating windows.}
  \label{fig:enviroments}
\end{figure}

\paragraph{Environments} We used two different 3D real-world scans as MR environments of our study: a coffee shop and a home office.
Figure \ref{fig:enviroments} illustrates them, adapted from the SemanticAdapt codebase \cite{cheng2021semanticadapt}.

In the coffee shop environment, we recreated a leisure scenario where participants were seated at a coffee table with physical items such as a smartphone, an iPad, or a cup of coffee.
The virtual UI elements provided included Instagram, a video viewer, a messenger, a photo viewer, two news web pages, and a music player.
Participants could personalize the placement of these UI elements either above the coffee table or in the empty space to their left.

In the home office environment, we designed a literature search scenario where participants were seated at a desk with objects like a phone, headphones, a book, and a tablet.
The interface elements included two papers, two web search engines, a messenger, a music player, and Slack. Participants could personalize the placement of interface elements above the desk or to their left or right. 

We intentionally selected and designed the scenarios to vary in visual complexity, available space, and social setting. 
These factors may influence users' mental models and lead to differences in their preferences.
The broader range of factors also strengthens our investigation into the approach's generalization.
We provided participants with task descriptions tailored to each specific scenario (see Appendix \ref{app:scenario-explanations}).

 

\paragraph{Methods} We compared three different \ivmethod{}s: \cmanual{}, \cauit{}, and \cours{}.
To ensure a fair comparison, all conditions started with the same initial layout randomly sampled from a Pareto frontier generated by NSGA-II. All conditions used the same set of objective functions (defined in Section \ref{sec:objective-terms}).

\begin{itemize}
    \item{ 
    \cmanual{}: Participants could move all elements of the initial design to create their preferred layout. No further layout optimization was conducted. Participants manually adjusted a layout until they were satisfied with it.
    }
    
    \item{ 
    \cauit{}: It allowed participants to also move elements as they wished, but instead of directly using these positions, we applied Step 3 of our approach to determine the Pareto-optimal layout closest to their adjustments (Section \ref{sec:determining-optimal-design}). 
    Participants could choose to move between one and three elements, after which the next Pareto-optimal layout was selected and displayed. 
    \rev{This cycle could be repeated as many times as needed until they were satisfied with the layout.}
    
    We designed \cauit{} to represent the state of the art in Pareto-optimization-based UI adaptation \cite{johns2023towards}.
    This approach suggests selecting a final design candidate from the Pareto frontier of an optimized single UI element by displaying different positions as "ghosts" for the user to choose from. 
    However, this method does not scale well to full UI layouts with multiple elements. 
    To address this, we implemented \cauit{} as a straightforward approach to integrate candidate selection of full UI layouts into non-prioritized MOO problems.
    By using the same candidate selection method as in our approach, we ensure that any differences arise solely from the contrast between prioritized and non-prioritized MOO problems.
    }

    \item{ 
    \cours{}: It represents our full approach described in Section~\ref{sec:Method}. 
    Based on participants' adjustments, their preference for objectives was determined, which was used to guide the prioritized Pareto frontier exploration. Then, the final optimal layout was identified.
    Again, participants could move between one and three elements, after which PL-NSGA-II recomputed the Pareto frontier. 
    \rev{This process could be repeated as many times as needed until they were satisfied with the layout.}
    }
\end{itemize}

\subsection{Procedure}
Prior to the study, participants completed a consent form and a pre-questionnaire to gather demographic data and assess their previous VR/MR experience. 
They were then given a tutorial on using hand gestures to interact with UI elements and had the opportunity to explore interactions and elements in a sample scene.
Participants proceeded to perform both free and targeted personalization tasks across the literature search and leisure environments, using each of the three adaptation methods. After completing each condition, they filled out a NASA-TLX questionnaire and ranked the layouts generated by each method.
At the end of the session, participants completed the UEQ-S and stated their method preference. Each session lasted approximately 30 minutes.

\definecolor{RankOneColor}{RGB}{78,154,6}  
\definecolor{RankTwoColor}{RGB}{208,124,64} 
\definecolor{RankThreeColor}{RGB}{67,99,216} 

\begin{figure*}[thb]
  \centering  
    \begin{subfigure}{0.32\textwidth}
        \centering
        \includegraphics[width=.95\linewidth]
        {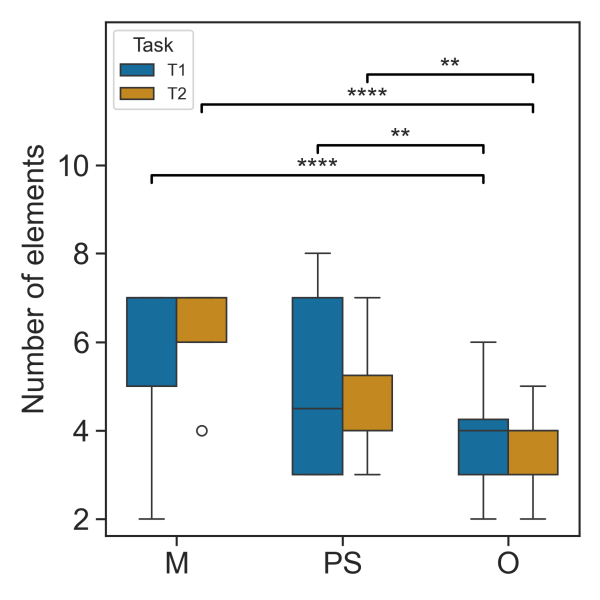}  
        \caption{}
    \end{subfigure}
    \begin{subfigure}{0.32\textwidth}
        \centering
        \includegraphics[width=.95\linewidth]
        {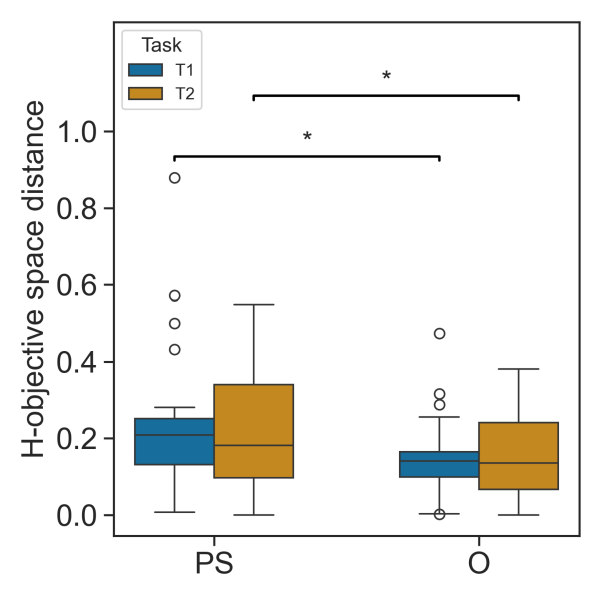}  
        \caption{}
    \end{subfigure}
    \begin{subfigure}{0.32\textwidth}
        \centering
        \includegraphics[width=.95\linewidth]
        {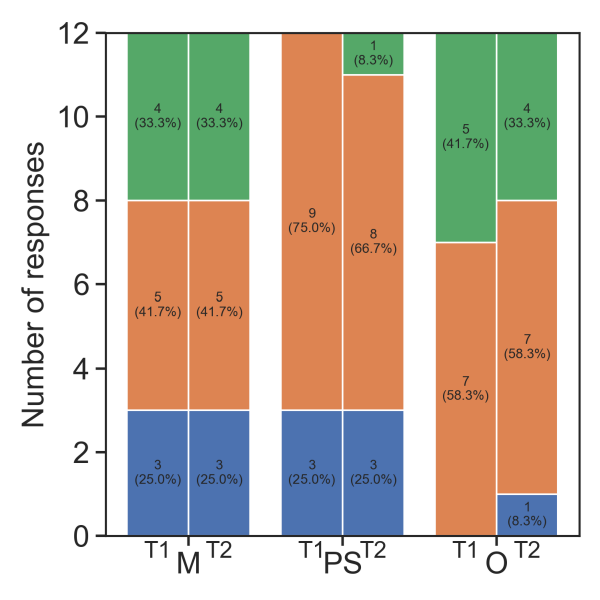}  
        \caption{}
    \end{subfigure}
  \caption{Results from our study of the three methods \rev{per task}: \cours{} (O), \cauit{} (PS), and \cmanual{} (M). The metrics displayed are (a) the number of moved elements, (b) H-objective space distance, and (c) layout preference. (a) and (b) are presented as boxplots, with significance indicated at the following levels: *p < .05, **p < .01, ***p < .001, ****p < .0001. (c) Illustrates the distribution of preference ranks among the three methods, categorized into {\color{RankOneColor} Rank 1}, {\color{RankTwoColor} Rank 2}, {\color{RankThreeColor} Rank 3}.}
  \Description{
  Box plots comparing three layout conditions: Manual (M), ParetoSelect (PS), and Ours (O). (a) Number of moved elements in each condition, with significantly less movements for Ours compared to ParetoSelect and Manual. (b) User layout preference ratings, with a preference for Manual and Ours. (c) H-objective space distance, with significant differences between the ParetoSelect and Ours conditions, with Ours being better.
  }
  \label{fig:result}
\end{figure*}

\subsection{Apparatus}
We extended our Unity application to include the baseline conditions and integrated the NSGA-II algorithm into our Python codebase.
Participants completed the study while seated, using an Oculus Quest 2 headset tethered to a computer via cable.
They used their hands to interact with UI elements via the Quest's hand tracking.

\subsection{Participants}
We recruited 12 participants (2 female, 10 male), aged 22--27 (Mean = 25, SD = 1.7) from a local university. Participants disclosed their familiarity with VR/MR technology.
Three participants considered themselves familiar but not experts, while one indicated no prior experience. The remaining participants reported occasional usage, possessing basic operational knowledge of VR devices. 

\section{Results}
\label{sec:results}
In the following, we present the results of our study.
For significance testing, we performed an ANOVA if data was normally distributed (Shapiro-Wilk \pvall{>}{.05}) and exhibited equal variances between groups (Levene's \pvall{>}{.05}).
Pairwise comparisons were performed using t-tests with Bonferroni-adjusted p-values. 
If either assumption was violated, 
we assessed differences with the non-parametric Aligned Rank Transform (ART) ANOVA. 
Post-hoc comparisons were then performed using the ART-C algorithm and Bonferroni corrections.
If not stated differently, \ivmethod{} and \ivtask{} were considered as within-subject factor and participants as a random factor.
\rev{Since statistical tests showed no significant differences between \ivtask{}s across any metrics, we focus solely on effects related to \ivmethod{} in the remainder of this section.}
Results with significant differences are shown in Figure \ref{fig:result}.

\subsection{Usability metrics}
We analyzed the collected metrics related to usability and user experience.
A main effect of \ivmethod{} on the \emph{number of moved elements} across \ivtask{}s \anova{2}{66}{18.97}{<}{.0001}{.38}~ was found.
Participants manually moved less UI elements with \cours{} compared to both \cauit{} \rev{\pvalr{=}{.01}{.35}} and \cmanual{} \rev{\pvalr{<}{.0001}{.61}}. 
No other significant differences were identified.

We evaluated the subjective feedback provided by the participants in-between conditions.
A significant effect of \ivmethod{} on \emph{layout preference} across \ivtask{}s was found \anova{2}{66}{3.98}{=}{.02}{.11}~.
Post-hoc tests have shown that participants ranked UI layouts generated by \cours{} higher than when generated by \cauit{} \rev{\pvalr{=}{.02}{.33}}.
Other differences were non-significant.
Additionally, no significant differences were identified for \emph{NASA-TLX}.

To statistically analyze the subjective feedback from the post-questionnaire, conditions were treated as a within-subject factor, while participants were treated as a random factor.
No significant differences were found for \emph{method preference} (\ivmethod{} \pvall{=}{.09}), nor for the dimensions of the UEQ: \emph{pragmatic quality} (\ivmethod{} \pvall{=}{.07}), \emph{hedonic quality} (\ivmethod{} \pvall{=}{.11}), and \emph{overall quality} (\ivmethod{} \pvall{=}{.07}).

\begin{figure*}[thb]
    \centering
    \begin{subfigure}{0.4\textwidth}
        \centering
        \includegraphics[width=.95\linewidth]
        {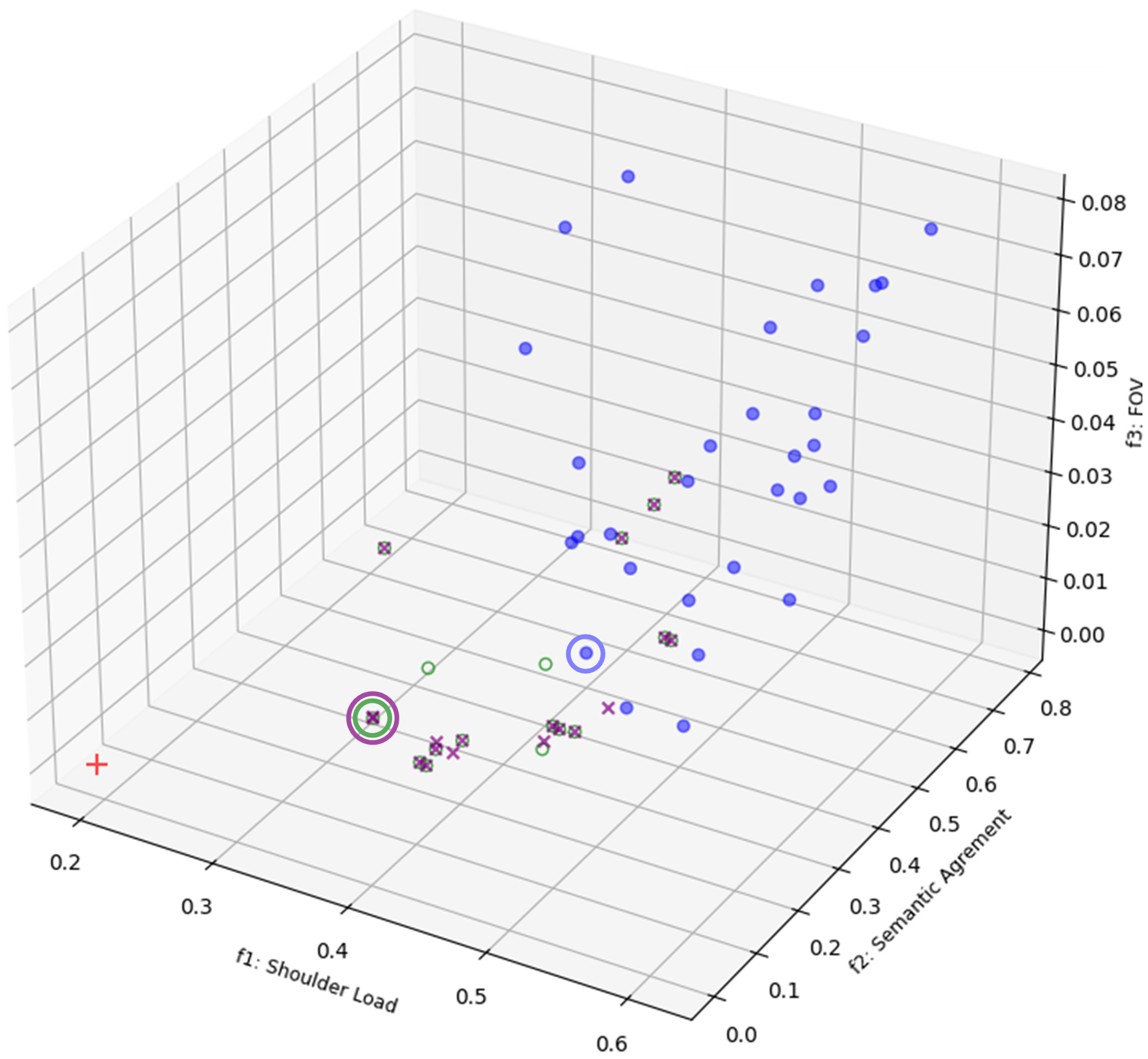}  
        \caption{}
    \end{subfigure}
    \begin{subfigure}{0.4\textwidth}
        \centering
        \includegraphics[width=.95\linewidth]
        {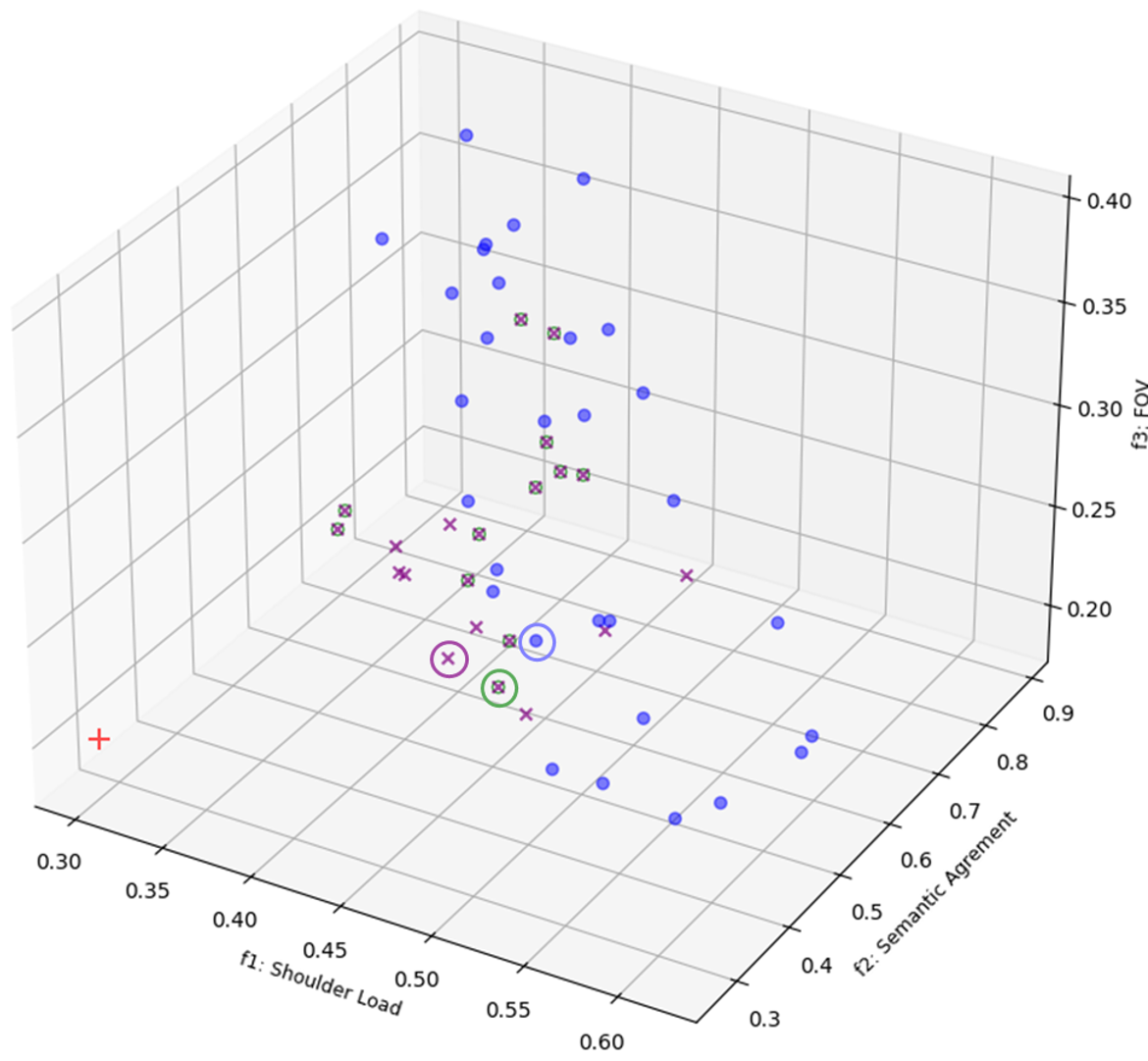}  
        \caption{}
    \end{subfigure}
    
    \caption{Pareto fronts of two study participants evolving through successive iterations of our algorithm. 
    The \protect\tikz \protect\draw[blue, fill=blue] (0,0) circle (0.08); represents the initial NSGA-II Pareto Front, 
    the \protect\tikz \protect\draw[green, thick] (0,0) circle (0.08); represents the PL-NSGA-II Pareto Front at iteration 1, the \protect\tikz \protect\node[purple] at (0,0) {\footnotesize $\times$}; represents the PL-NSGA-II Pareto Front at iteration 2, and the \protect\tikz \protect\node[red] at (0,0) {\footnotesize $+$}; represents the reference point attained from all user moved UI elements. 
    The larger circles (\protect\tikz \protect\draw[blue, thick] (0,0) circle (0.1);, \protect\tikz \protect\draw[green, thick] (0,0) circle (0.1);, \protect\tikz \protect\draw[purple, thick] (0,0) circle (0.1);) surrounding the Pareto front candidates represent the best candidates from the algorithm iterations with corresponding colors.
    }
    
    \Description[]{

    A side-by-side comparison of two 3D scatter plots labeled (a) and (b), displaying Pareto fronts for two study participants. The plots depict three objective functions: f1 (Shoulder Load), f2 (Semantic Alignment), and f3 (FOV). Data points are represented with different markers and colors: small blue circles for general Pareto front candidates, green, purple, and blue circles indicating the best candidates in their respective algorithm iterations, purple "x" markers for intermediate candidates, and red "+" symbols marking reference points. The distributions of points vary slightly between the two participants, reflecting their progression through algorithm iterations.
    
    }
    \label{fig:pareto_front}
\end{figure*}

\subsection{Optimization metrics}
We evaluated the metrics computed from the optimization problems themselves to further compare \cauit{} and \cours{}. 
No significant differences were found in the \emph{hypervolume} and the \emph{average costs} of objective functions within all groups ($H,M,L$) as well as within priority group $H$.

Since the \emph{distance in objective space} of moved UI elements to the nearest candidate on the Pareto front depends on UI element movement, we must ensure these movements are the same for both \cauit{} and \cours{} for a fair comparison. 
Therefore, for each condition, we recomputed the Pareto front using the movements from participants in the other condition. 
To track which condition the movements came from and assess a potential effect, we introduced the factor \ivcomp{}. 
For statistical analysis, \ivmethod{} (the method that generates the Pareto-front designs) was treated as a within-subject factor, and \ivcomp{} (the condition where the user adjustment was originally from)  as a between-subject factor.

No differences between conditions were identified for the \emph{distance in objective space} of all objective functions.
However, we found a main effect of \ivmethod{} on the \emph{objective space distance} of high priority objectives ($H$) across \ivcomp{}s \anova{1}{92}{4.26}{=}{.04}{.04}~.
No other significant differences were identified.

The advantage of our method in achieving a lower distance in the objective space, becomes evident when analyzing the temporal progression of the Pareto front across iterations of our algorithm.
Figure 5 illustrates this progression for two participants in our study, demonstrating that the best solution candidates of our PL-NSGA-II algorithm converge closer to the reference point derived from user-moved UI elements. Moreover, these candidates consistently dominate those from the Pareto front generated by NSGA-II.

\section{Discussion}
\label{sec:discussion}

The goal of our approach is to improve multi-objective UI optimization by more effectively incorporating users' individual preferences in the adaptation of UI layouts.
Our generated layouts should require fewer adjustments from users when customizing them to their needs, while increasing overall satisfaction with the layout.

Our results indicate that these objectives were achieved.
With \cours{}, participants required fewer UI element movements to customize layouts to their preferences compared to both \cauit{} and \cmanual{}. 
Additionally, participants expressed a clear preference for layouts generated by \cours{} over those by \cauit{}.
No significant difference between \cours{} and \cmanual{}, where users could fully control the layout, suggests that \cours{} maintains the flexibility of manual customization while reducing its effort (less adjustments).
\rev{Note that statistical tests showed no differences between \ivtask{}s across any metrics. These results suggest that our method’s advantages hold for cases when objectives are known (Task 2) to more realistic scenarios where users are unfamiliar with them (Task 1).}

\rev{The higher preference for layouts generated by our method did not yet translate into a clear overall preference or higher user experience compared to the baselines.
This is not surprising, as UX metrics such as NASA-TLX and UEQ-S are known to be insensitive to short-term UI adaptation tasks \cite{10.1145/3613904.3642071,lee2020autogain}. Nevertheless, the results of our UX-related measures indicate promising trends.}
Notably, the p-values for the related metrics were relatively low \pval{=}{.07~-~.11}, with \cours{} consistently trending better than baselines across criteria. 
This suggests that with a larger sample size, the results have strong potential to achieve statistical significance.

We further explored whether the performance in these usability-related metrics can be attributed to the underlying algorithm's functionality.
Results showed that \cours{} achieved a lower distance in objective space between moved UI elements and the nearest candidate on the Pareto front for the group of objectives with highest priority, compared to \cauit{}. 
%
This shows that \cours{} resulted in Pareto frontiers that are more focused and aligned with the users' prioritized objectives than other baselines.
Furthermore, we highlight that while \cours{} generates a more focused, \rev{user-aligned} search (reflected by the shorter distances for high-priority objectives), it maintains overall design space exploration, as indicated by the equivalent hypervolumes across \ivmethod{}.
\rev{These findings indicate that our method directs the search more effectively toward regions of the Pareto frontier that better reflect user intent, all within the same search budget.}

\rev{A known limitation of UI optimization is that some user preferences may lie outside the defined objectives. As with state-of-the-art approaches, we assume that the selected objectives capture the majority of user needs. The validity of this assumption is supported by the positive results in our free personalization task, where our method still reduced widget adjustments and led to a preference for the generated layout, highlighting the relevance of the chosen objective set. To mitigate potential mismatches between modeled objectives and user preferences, our approach (a) downranks irrelevant objectives via lexicographic inference and (b) allows manual adjustments beyond the modeled space.}

\section{Limitations \& Future Work}

\paragraph{Hard-constraint modeling of moved UI elements}

In our current approach, user-adjusted UI elements are leveraged to estimate objective preferences and subsequently identify the optimal UI layout on the Pareto front. 
As a result, elements manually moved by the user may still be adjusted by the algorithm in later iterations. 
This is an intentional design choice, as it allows the algorithm to optimize all elements in an unconstrained space with increasing information on user preference.
However, future work could explore the integration of user-adjusted UI elements as hard constraints within the optimization process and investigate whether this enhances the perceived usefulness of our approach.

\paragraph{Handling uncertainty in user preferences}

Our approach infers user preferences in a deterministic manner, assuming that preferences can be directly derived from changes in objective values. However, in real-world scenarios, user adjustments may be inconsistent, exploratory, or even accidental, introducing noise into the preference estimation process. While we have applied filtering techniques to mitigate this issue, more advanced approaches should be explored in future work.
A promising direction is to move beyond direct value-based inference and instead adopt probabilistic methods, such as Bayesian preference modeling or probabilistic ranking, which can incorporate confidence levels in inferred preferences. Additionally, stochastic optimization techniques could be leveraged to dynamically refine preference estimations while accounting for uncertainty and inconsistency in user interactions. These approaches would enhance robustness, making UI adaptation more resilient to noisy or ambiguous user feedback.

\paragraph{Potential of dynamic Pareto frontier exploration}
In this paper, we employ PL-NSGA-II, which ranks design objectives to guide the dynamic exploration of the Pareto frontier. 
Our approach outperformed traditional NSGA, whose Pareto front exploration lacks guidance from additional information. This demonstrates the potential of dynamically informed Pareto-optimal exploration.
Future work should investigate other relevant approaches for such dynamic exploration. 
One straightforward yet under-explored approach in HCI is the adaptive weighted-sum method \cite{kim2006adaptive}, which actively learns user preferences for different objective weightings and converges toward a preference-guided optimum.
A more complex direction is Trust-Region Optimization \cite{carrizo2016trust,mohammadi2024trust}, where the search process is focused within a ``trust region'' or around a ``trust point'' in the design space or objective space, prioritizing exploration in this local area. 
Trust-region methods have been adapted for evolutionary algorithms, e.g., Reference Point-Based NSGA-II \cite{deb2006reference}, which could be naturally extended to our case. 
The user adjustment could provide the trust region, enabling a more focused search around nearby designs.
Another promising direction is incorporating surrogate models (e.g., Gaussian Processes) to learn user preferences based on their interaction, and guide exploration dynamically \cite{moon2023amortized}.
Future research is encouraged to explore these dynamic Pareto-frontier exploration techniques and assess their potential benefits and challenges in various design contexts.

\paragraph{Beyond the preferences in objective space} 

Our approach assumes that users have implicit preferences over objectives, yet, in practice, their preferences may also exist in the design space. For example, a user might prefer certain UI elements to remain fixed in specific locations, even if these preferences do not align with the predefined design objectives.
Preference optimization is a parallel topic focused on learning and optimizing user preferences within the design space \cite{yamamoto2022photographic}. Exploring a hybrid approach that infers whether the user's preference lies within the objective space (e.g., enhancing overall reachability) or the design space (e.g., positioning certain elements in specific locations) would be a valuable direction for future research. By understanding these distinctions, the optimizer can adapt its exploration strategy accordingly, leading to more personalized and effective design solutions.

\paragraph{Open challenges in re-optimization}
\rev{Consistent with prior UI optimization work \cite{bailly2013menuoptimizer,cheng2021semanticadapt,cheng2023interactionadapt}, we assume a one-time adaptation based on stable user preferences where users can restart the process if their preferences change.}
In practice, nonetheless, user preferences may evolve over time, which calls for the capability to dynamically re-adapt. 
This introduces several open research questions. For instance, how can we accurately detect when a user’s preferences have significantly shifted? How frequently should the system re-adapt? How can the system re-adapt while balancing the preservation of the current design to avoid drastic changes?
Future research should consider addressing these challenges to create systems capable of long-term, dynamic interaction and evolving user satisfaction. 
Further, future works should extend the optimization to a continual setting \cite{10.1145/3706598.3713603}; that is, the system keeps accumulating past experience and becomes more informed to make better decisions over time.

\paragraph{Transferring learned preferences across UIs and users}

Currently, our approach determines user preferences from scratch for each UI adaptation, requiring users to manually adjust layouts even when their preferences remain consistent across similar interfaces. This redundancy increases user effort and slows down optimization. Future work should explore meta-learning or transfer learning to carry over learned preferences from one UI to another, particularly when interfaces share structural similarities. By leveraging past interactions, the system could provide more informed initial suggestions, reducing the need for repeated manual refinements.

Beyond individual users, preference learning could also be extended across users interacting with the same or similar UIs. If multiple users have optimized an interface, new users could benefit from these collective insights, receiving pre-optimized layouts that align with commonly preferred patterns. Clustering techniques could help group users with similar preferences, enabling a personalized yet scalable adaptation strategy. Future research should explore how to balance shared knowledge transfer with individual customization, ensuring UI layouts remain flexible while benefiting from prior optimizations.

\paragraph{Objective function parameters} 
Parameters like arm's length, ideal viewing distance from an interface, and levels of semantic association and avoidance impact the position of minimum and maximum points in the objective functions. 
Currently, population averages are used to set these values, but fixed parameters cannot account for individual variations and needs. 
To further tailor MR layouts to individual users' preferences and needs, future work should explore the feasibility of automatically inferring these parameters from users' interaction behavior or other sources.

\paragraph{Scalability}
\rev{
In this work, we demonstrated that our approach performs effectively with five objectives and seven widgets. However, real-world use cases may involve a greater number of objectives and widgets, potentially leading to longer optimization runtimes. To address this, we employ NSGA-II \cite{deb2002fast, johns2023towards}, a widely used multi-objective optimizer known for its linear runtime scaling with respect to the number of input elements and objectives. Combined with its support for parallelization and a fixed search budget, it enables scalable and predictable real-time performance.}

\section{Conclusion}
In this paper, we introduced a novel preference-guided multi objective optimization approach for UI layout adaptation, addressing key limitations of traditional MOO methods. 
By utilizing mixed Pareto-lexicographic optimization, our approach dynamically explores the Pareto frontier based on derived user preference rankings over objective terms, ensuring that the resulting designs are better aligned with the user's needs. 
This enables more personalized UI adaptation in both 2D interfaces and 3D Mixed Reality environments.

Through a user study in an MR context, we demonstrated that our method significantly reduces the effort required for manual UI adjustments while maintaining high user satisfaction with the final layout. These findings highlight the potential of preference-guided optimization in improving the effectiveness of adaptive UIs.

Overall, our contributions provide a strong foundation towards the true personalization of adaptive UIs.
The application of mixed Pareto-lexicographic MOO presents a promising direction for future UI adaptation frameworks, for instance, enabling hybrid approaches that combine it with other methods to directly optimize for user preferences within the design space.

\begin{acks}
Yi-Chi Liao was supported by the ETH Zurich Postdoctoral Fellowship Programme.
\end{acks}
\bibliographystyle{ACM-Reference-Format}
\bibliography{references}


\begin{thebibliography}{33}


\ifx \showCODEN    \undefined \def \showCODEN     #1{\unskip}     \fi
\ifx \showDOI      \undefined \def \showDOI       #1{#1}\fi
\ifx \showISBNx    \undefined \def \showISBNx     #1{\unskip}     \fi
\ifx \showISBNxiii \undefined \def \showISBNxiii  #1{\unskip}     \fi
\ifx \showISSN     \undefined \def \showISSN      #1{\unskip}     \fi
\ifx \showLCCN     \undefined \def \showLCCN      #1{\unskip}     \fi
\ifx \shownote     \undefined \def \shownote      #1{#1}          \fi
\ifx \showarticletitle \undefined \def \showarticletitle #1{#1}   \fi
\ifx \showURL      \undefined \def \showURL       {\relax}        \fi
\providecommand\bibfield[2]{#2}
\providecommand\bibinfo[2]{#2}
\providecommand\natexlab[1]{#1}
\providecommand\showeprint[2][]{arXiv:#2}

\bibitem[Ankrum and Nemeth(1995)]%
        {ankrum1995posture}
\bibfield{author}{\bibinfo{person}{Dennis~R Ankrum} {and} \bibinfo{person}{Kristie~J Nemeth}.} \bibinfo{year}{1995}\natexlab{}.
\newblock \showarticletitle{Posture, comfort, and monitor placement}.
\newblock \bibinfo{journal}{\emph{Ergonomics in design}} \bibinfo{volume}{3}, \bibinfo{number}{2} (\bibinfo{year}{1995}), \bibinfo{pages}{7--9}.
\newblock
\urldef\tempurl%
\url{https://doi.org/10.1177/106480469500300203}
\showURL{%
\tempurl}


\bibitem[Bailly et~al\mbox{.}(2013)]%
        {bailly2013menuoptimizer}
\bibfield{author}{\bibinfo{person}{Gilles Bailly}, \bibinfo{person}{Antti Oulasvirta}, \bibinfo{person}{Timo K{\"o}tzing}, {and} \bibinfo{person}{Sabrina Hoppe}.} \bibinfo{year}{2013}\natexlab{}.
\newblock \showarticletitle{Menuoptimizer: Interactive optimization of menu systems}. In \bibinfo{booktitle}{\emph{Proceedings of the 26th annual ACM symposium on User interface software and technology}}. \bibinfo{pages}{331--342}.
\newblock
\urldef\tempurl%
\url{https://doi.org/10.1145/2501988.2502024}
\showURL{%
\tempurl}


\bibitem[Blank and Deb(2020)]%
        {blank2020pymoo}
\bibfield{author}{\bibinfo{person}{Julian Blank} {and} \bibinfo{person}{Kalyanmoy Deb}.} \bibinfo{year}{2020}\natexlab{}.
\newblock \showarticletitle{Pymoo: Multi-objective optimization in python}.
\newblock \bibinfo{journal}{\emph{Ieee access}}  \bibinfo{volume}{8} (\bibinfo{year}{2020}), \bibinfo{pages}{89497--89509}.
\newblock


\bibitem[Carrizo et~al\mbox{.}(2016)]%
        {carrizo2016trust}
\bibfield{author}{\bibinfo{person}{Gabriel~A Carrizo}, \bibinfo{person}{Pablo~A Lotito}, {and} \bibinfo{person}{Mar{\'\i}a~C Maciel}.} \bibinfo{year}{2016}\natexlab{}.
\newblock \showarticletitle{Trust region globalization strategy for the nonconvex unconstrained multiobjective optimization problem}.
\newblock \bibinfo{journal}{\emph{Mathematical Programming}}  \bibinfo{volume}{159} (\bibinfo{year}{2016}), \bibinfo{pages}{339--369}.
\newblock


\bibitem[Chan et~al\mbox{.}(2022)]%
        {chan2022investigating}
\bibfield{author}{\bibinfo{person}{Liwei Chan}, \bibinfo{person}{Yi-Chi Liao}, \bibinfo{person}{George~B Mo}, \bibinfo{person}{John~J Dudley}, \bibinfo{person}{Chun-Lien Cheng}, \bibinfo{person}{Per~Ola Kristensson}, {and} \bibinfo{person}{Antti Oulasvirta}.} \bibinfo{year}{2022}\natexlab{}.
\newblock \showarticletitle{Investigating positive and negative qualities of human-in-the-loop optimization for designing interaction techniques}. In \bibinfo{booktitle}{\emph{Proceedings of the 2022 CHI Conference on Human Factors in Computing Systems}}. \bibinfo{pages}{1--14}.
\newblock
\urldef\tempurl%
\url{https://doi.org/10.1145/3491102.3501850}
\showURL{%
\tempurl}


\bibitem[Cheng et~al\mbox{.}(2021)]%
        {cheng2021semanticadapt}
\bibfield{author}{\bibinfo{person}{Yifei Cheng}, \bibinfo{person}{Yukang Yan}, \bibinfo{person}{Xin Yi}, \bibinfo{person}{Yuanchun Shi}, {and} \bibinfo{person}{David Lindlbauer}.} \bibinfo{year}{2021}\natexlab{}.
\newblock \showarticletitle{Semanticadapt: Optimization-based adaptation of mixed reality layouts leveraging virtual-physical semantic connections}. In \bibinfo{booktitle}{\emph{The 34th Annual ACM Symposium on User Interface Software and Technology}}. \bibinfo{pages}{282--297}.
\newblock
\urldef\tempurl%
\url{https://doi.org/10.1145/3472749.3474750}
\showURL{%
\tempurl}


\bibitem[Cheng et~al\mbox{.}(2023)]%
        {cheng2023interactionadapt}
\bibfield{author}{\bibinfo{person}{Yi~Fei Cheng}, \bibinfo{person}{Christoph Gebhardt}, {and} \bibinfo{person}{Christian Holz}.} \bibinfo{year}{2023}\natexlab{}.
\newblock \showarticletitle{InteractionAdapt: Interaction-driven Workspace Adaptation for Situated Virtual Reality Environments}. In \bibinfo{booktitle}{\emph{Proceedings of the 36th Annual ACM Symposium on User Interface Software and Technology}}. \bibinfo{pages}{1--14}.
\newblock
\urldef\tempurl%
\url{https://doi.org/10.1145/3586183.3606717}
\showURL{%
\tempurl}


\bibitem[Deb et~al\mbox{.}(2002)]%
        {deb2002fast}
\bibfield{author}{\bibinfo{person}{Kalyanmoy Deb}, \bibinfo{person}{Amrit Pratap}, \bibinfo{person}{Sameer Agarwal}, {and} \bibinfo{person}{TAMT Meyarivan}.} \bibinfo{year}{2002}\natexlab{}.
\newblock \showarticletitle{A fast and elitist multiobjective genetic algorithm: NSGA-II}.
\newblock \bibinfo{journal}{\emph{IEEE transactions on evolutionary computation}} \bibinfo{volume}{6}, \bibinfo{number}{2} (\bibinfo{year}{2002}), \bibinfo{pages}{182--197}.
\newblock


\bibitem[Deb and Sundar(2006)]%
        {deb2006reference}
\bibfield{author}{\bibinfo{person}{Kalyanmoy Deb} {and} \bibinfo{person}{Jayavelmurugan Sundar}.} \bibinfo{year}{2006}\natexlab{}.
\newblock \showarticletitle{Reference point based multi-objective optimization using evolutionary algorithms}. In \bibinfo{booktitle}{\emph{Proceedings of the 8th annual conference on Genetic and evolutionary computation}}. \bibinfo{pages}{635--642}.
\newblock


\bibitem[Diego-Mas et~al\mbox{.}(2019)]%
        {uioptimizationgenetic2019diegomas}
\bibfield{author}{\bibinfo{person}{Jose~Antonio Diego-Mas}, \bibinfo{person}{Diana Garzon-Leal}, \bibinfo{person}{Rocio Poveda-Bautista}, {and} \bibinfo{person}{Jorge Alcaide-Marzal}.} \bibinfo{year}{2019}\natexlab{}.
\newblock \showarticletitle{User-interfaces layout optimization using eye-tracking, mouse movements and genetic algorithms}.
\newblock \bibinfo{journal}{\emph{Applied Ergonomics}}  \bibinfo{volume}{78} (\bibinfo{year}{2019}), \bibinfo{pages}{197--209}.
\newblock
\showISSN{0003-6870}
\urldef\tempurl%
\url{https://doi.org/10.1016/j.apergo.2019.03.004}
\showDOI{\tempurl}


\bibitem[Evangelista~Belo et~al\mbox{.}(2022)]%
        {evangelista2022auit}
\bibfield{author}{\bibinfo{person}{Jo{\~a}o~Marcelo Evangelista~Belo}, \bibinfo{person}{Mathias~N Lystb{\ae}k}, \bibinfo{person}{Anna~Maria Feit}, \bibinfo{person}{Ken Pfeuffer}, \bibinfo{person}{Peter K{\'a}n}, \bibinfo{person}{Antti Oulasvirta}, {and} \bibinfo{person}{Kaj Gr{\o}nb{\ae}k}.} \bibinfo{year}{2022}\natexlab{}.
\newblock \showarticletitle{Auit--the adaptive user interfaces toolkit for designing xr applications}. In \bibinfo{booktitle}{\emph{Proceedings of the 35th Annual ACM Symposium on User Interface Software and Technology}}. \bibinfo{pages}{1--16}.
\newblock
\urldef\tempurl%
\url{https://doi.org/10.1145/3526113.3545651}
\showURL{%
\tempurl}


\bibitem[Gunantara(2018)]%
        {gunantara2018review}
\bibfield{author}{\bibinfo{person}{Nyoman Gunantara}.} \bibinfo{year}{2018}\natexlab{}.
\newblock \showarticletitle{A review of multi-objective optimization: Methods and its applications}.
\newblock \bibinfo{journal}{\emph{Cogent Engineering}} \bibinfo{volume}{5}, \bibinfo{number}{1} (\bibinfo{year}{2018}), \bibinfo{pages}{1502242}.
\newblock


\bibitem[Hart(1986)]%
        {Hart1986}
\bibfield{author}{\bibinfo{person}{Sandra~G Hart}.} \bibinfo{year}{1986}\natexlab{}.
\newblock \showarticletitle{NASA task load index (TLX)}.
\newblock \bibinfo{journal}{\emph{NASA Ames Research Center}} (\bibinfo{year}{1986}).
\newblock
\urldef\tempurl%
\url{https://ntrs.nasa.gov/citations/20000021487}
\showURL{%
\tempurl}


\bibitem[Hofmann et~al\mbox{.}(2023)]%
        {hofmann2023optimism}
\bibfield{author}{\bibinfo{person}{Megan Hofmann}, \bibinfo{person}{Nayha Auradkar}, \bibinfo{person}{Jessica Birchfield}, \bibinfo{person}{Jerry Cao}, \bibinfo{person}{Autumn~G Hughes}, \bibinfo{person}{Gene~SH Kim}, \bibinfo{person}{Shriya Kurpad}, \bibinfo{person}{Kathryn~J Lum}, \bibinfo{person}{Kelly Mack}, \bibinfo{person}{Anisha Nilakantan}, {et~al\mbox{.}}} \bibinfo{year}{2023}\natexlab{}.
\newblock \showarticletitle{OPTIMISM: Enabling Collaborative Implementation of Domain Specific Metaheuristic Optimization}. In \bibinfo{booktitle}{\emph{Proceedings of the 2023 CHI Conference on Human Factors in Computing Systems}}. \bibinfo{pages}{1--19}.
\newblock
\urldef\tempurl%
\url{https://doi.org/10.1145/3544548.3580904}
\showURL{%
\tempurl}


\bibitem[Johns et~al\mbox{.}(2023)]%
        {johns2023towards}
\bibfield{author}{\bibinfo{person}{Christoph~Albert Johns}, \bibinfo{person}{Jo{\~a}o~Marcelo Evangelista~Belo}, \bibinfo{person}{Anna~Maria Feit}, \bibinfo{person}{Clemens~Nylandsted Klokmose}, {and} \bibinfo{person}{Ken Pfeuffer}.} \bibinfo{year}{2023}\natexlab{}.
\newblock \showarticletitle{Towards Flexible and Robust User Interface Adaptations With Multiple Objectives}. In \bibinfo{booktitle}{\emph{Proceedings of the 36th Annual ACM Symposium on User Interface Software and Technology}}. \bibinfo{pages}{1--17}.
\newblock
\urldef\tempurl%
\url{https://doi.org/10.1145/3586183.3606799}
\showURL{%
\tempurl}


\bibitem[Kim and de~Weck(2006)]%
        {kim2006adaptive}
\bibfield{author}{\bibinfo{person}{Il~Yong Kim} {and} \bibinfo{person}{Olivier~L de Weck}.} \bibinfo{year}{2006}\natexlab{}.
\newblock \showarticletitle{Adaptive weighted sum method for multiobjective optimization: a new method for Pareto front generation}.
\newblock \bibinfo{journal}{\emph{Structural and multidisciplinary optimization}} \bibinfo{volume}{31}, \bibinfo{number}{2} (\bibinfo{year}{2006}), \bibinfo{pages}{105--116}.
\newblock


\bibitem[Lai et~al\mbox{.}(2020)]%
        {lai2020solving}
\bibfield{author}{\bibinfo{person}{Leonardo Lai}, \bibinfo{person}{Lorenzo Fiaschi}, {and} \bibinfo{person}{Marco Cococcioni}.} \bibinfo{year}{2020}\natexlab{}.
\newblock \showarticletitle{Solving mixed Pareto-Lexicographic multi-objective optimization problems: The case of priority chains}.
\newblock \bibinfo{journal}{\emph{Swarm and Evolutionary Computation}}  \bibinfo{volume}{55} (\bibinfo{year}{2020}), \bibinfo{pages}{100687}.
\newblock


\bibitem[Lai et~al\mbox{.}(2021)]%
        {lai2021solving}
\bibfield{author}{\bibinfo{person}{Leonardo Lai}, \bibinfo{person}{Lorenzo Fiaschi}, \bibinfo{person}{Marco Cococcioni}, {and} \bibinfo{person}{Kalyanmoy Deb}.} \bibinfo{year}{2021}\natexlab{}.
\newblock \showarticletitle{Solving mixed pareto-lexicographic multiobjective optimization problems: the case of priority levels}.
\newblock \bibinfo{journal}{\emph{IEEE Transactions on Evolutionary Computation}} \bibinfo{volume}{25}, \bibinfo{number}{5} (\bibinfo{year}{2021}), \bibinfo{pages}{971--985}.
\newblock


\bibitem[Lee et~al\mbox{.}(2020)]%
        {lee2020autogain}
\bibfield{author}{\bibinfo{person}{Byungjoo Lee}, \bibinfo{person}{Mathieu Nancel}, \bibinfo{person}{Sunjun Kim}, {and} \bibinfo{person}{Antti Oulasvirta}.} \bibinfo{year}{2020}\natexlab{}.
\newblock \showarticletitle{AutoGain: Gain function adaptation with submovement efficiency optimization}. In \bibinfo{booktitle}{\emph{Proceedings of the 2020 CHI Conference on Human Factors in Computing Systems}}.
\newblock


\bibitem[Li et~al\mbox{.}(2024)]%
        {li2024situationadapt}
\bibfield{author}{\bibinfo{person}{Zhipeng Li}, \bibinfo{person}{Christoph Gebhardt}, \bibinfo{person}{Yves Inglin}, \bibinfo{person}{Nicolas Steck}, \bibinfo{person}{Paul Streli}, {and} \bibinfo{person}{Christian Holz}.} \bibinfo{year}{2024}\natexlab{}.
\newblock \showarticletitle{Situationadapt: Contextual ui optimization in mixed reality with situation awareness via llm reasoning}. In \bibinfo{booktitle}{\emph{Proceedings of the 37th Annual ACM Symposium on User Interface Software and Technology}}. \bibinfo{pages}{1--13}.
\newblock


\bibitem[Liao et~al\mbox{.}(2024)]%
        {10.1145/3613904.3642071}
\bibfield{author}{\bibinfo{person}{Yi-Chi Liao}, \bibinfo{person}{Ruta Desai}, \bibinfo{person}{Alec~M Pierce}, \bibinfo{person}{Krista~E. Taylor}, \bibinfo{person}{Hrvoje Benko}, \bibinfo{person}{Tanya~R. Jonker}, {and} \bibinfo{person}{Aakar Gupta}.} \bibinfo{year}{2024}\natexlab{}.
\newblock \showarticletitle{A Meta-Bayesian Approach for Rapid Online Parametric Optimization for Wrist-based Interactions}. In \bibinfo{booktitle}{\emph{Proceedings of the CHI Conference on Human Factors in Computing Systems}} (Honolulu, HI, USA) \emph{(\bibinfo{series}{CHI '24})}. \bibinfo{publisher}{Association for Computing Machinery}, \bibinfo{address}{New York, NY, USA}, Article \bibinfo{articleno}{410}, \bibinfo{numpages}{38}~pages.
\newblock
\showISBNx{9798400703300}
\urldef\tempurl%
\url{https://doi.org/10.1145/3613904.3642071}
\showDOI{\tempurl}


\bibitem[Liao et~al\mbox{.}(2023)]%
        {liao2023interaction}
\bibfield{author}{\bibinfo{person}{Yi-Chi Liao}, \bibinfo{person}{John~J Dudley}, \bibinfo{person}{George~B Mo}, \bibinfo{person}{Chun-Lien Cheng}, \bibinfo{person}{Liwei Chan}, \bibinfo{person}{Antti Oulasvirta}, {and} \bibinfo{person}{Per~Ola Kristensson}.} \bibinfo{year}{2023}\natexlab{}.
\newblock \showarticletitle{Interaction Design With Multi-objective Bayesian Optimization}.
\newblock \bibinfo{journal}{\emph{IEEE Pervasive Computing}} \bibinfo{volume}{22}, \bibinfo{number}{1} (\bibinfo{year}{2023}), \bibinfo{pages}{29--38}.
\newblock
\urldef\tempurl%
\url{https://doi.org/10.1109/MPRV.2022.3230597}
\showURL{%
\tempurl}


\bibitem[Liao et~al\mbox{.}(2025)]%
        {10.1145/3706598.3713603}
\bibfield{author}{\bibinfo{person}{Yi-Chi Liao}, \bibinfo{person}{Paul Streli}, \bibinfo{person}{Zhipeng Li}, \bibinfo{person}{Christoph Gebhardt}, {and} \bibinfo{person}{Christian Holz}.} \bibinfo{year}{2025}\natexlab{}.
\newblock \showarticletitle{Continual Human-in-the-Loop Optimization}. In \bibinfo{booktitle}{\emph{Proceedings of the 2025 CHI Conference on Human Factors in Computing Systems}} \emph{(\bibinfo{series}{CHI '25})}. \bibinfo{publisher}{Association for Computing Machinery}, \bibinfo{address}{New York, NY, USA}, Article \bibinfo{articleno}{795}, \bibinfo{numpages}{26}~pages.
\newblock
\showISBNx{9798400713941}
\urldef\tempurl%
\url{https://doi.org/10.1145/3706598.3713603}
\showDOI{\tempurl}


\bibitem[Marler and Arora(2004)]%
        {marler2004survey}
\bibfield{author}{\bibinfo{person}{R~Timothy Marler} {and} \bibinfo{person}{Jasbir~S Arora}.} \bibinfo{year}{2004}\natexlab{}.
\newblock \showarticletitle{Survey of multi-objective optimization methods for engineering}.
\newblock \bibinfo{journal}{\emph{Structural and multidisciplinary optimization}}  \bibinfo{volume}{26} (\bibinfo{year}{2004}), \bibinfo{pages}{369--395}.
\newblock
\urldef\tempurl%
\url{https://doi.org/10.1007/s00158-003-0368-6}
\showURL{%
\tempurl}


\bibitem[Miller(1995)]%
        {miller1995wordnet}
\bibfield{author}{\bibinfo{person}{George~A Miller}.} \bibinfo{year}{1995}\natexlab{}.
\newblock \showarticletitle{WordNet: a lexical database for English}.
\newblock \bibinfo{journal}{\emph{Commun. ACM}} \bibinfo{volume}{38}, \bibinfo{number}{11} (\bibinfo{year}{1995}), \bibinfo{pages}{39--41}.
\newblock
\urldef\tempurl%
\url{https://doi.org/10.1145/219717.219748}
\showURL{%
\tempurl}


\bibitem[Mohammadi and Cust{\'o}dio(2024)]%
        {mohammadi2024trust}
\bibfield{author}{\bibinfo{person}{A Mohammadi} {and} \bibinfo{person}{Ana~Lu{\'\i}sa Cust{\'o}dio}.} \bibinfo{year}{2024}\natexlab{}.
\newblock \showarticletitle{A trust-region approach for computing Pareto fronts in multiobjective optimization}.
\newblock \bibinfo{journal}{\emph{Computational Optimization and Applications}} \bibinfo{volume}{87}, \bibinfo{number}{1} (\bibinfo{year}{2024}), \bibinfo{pages}{149--179}.
\newblock


\bibitem[Moon et~al\mbox{.}(2023)]%
        {moon2023amortized}
\bibfield{author}{\bibinfo{person}{Hee-Seung Moon}, \bibinfo{person}{Antti Oulasvirta}, {and} \bibinfo{person}{Byungjoo Lee}.} \bibinfo{year}{2023}\natexlab{}.
\newblock \showarticletitle{Amortized inference with user simulations}. In \bibinfo{booktitle}{\emph{Proceedings of the 2023 CHI Conference on Human Factors in Computing Systems}}. \bibinfo{pages}{1--20}.
\newblock


\bibitem[Oulasvirta et~al\mbox{.}(2020)]%
        {oulasvirta2020combinatorial}
\bibfield{author}{\bibinfo{person}{Antti Oulasvirta}, \bibinfo{person}{Niraj~Ramesh Dayama}, \bibinfo{person}{Morteza Shiripour}, \bibinfo{person}{Maximilian John}, {and} \bibinfo{person}{Andreas Karrenbauer}.} \bibinfo{year}{2020}\natexlab{}.
\newblock \showarticletitle{Combinatorial optimization of graphical user interface designs}.
\newblock \bibinfo{journal}{\emph{Proc. IEEE}} \bibinfo{volume}{108}, \bibinfo{number}{3} (\bibinfo{year}{2020}), \bibinfo{pages}{434--464}.
\newblock
\urldef\tempurl%
\url{https://doi.org/10.1109/JPROC.2020.2969687}
\showURL{%
\tempurl}


\bibitem[Park et~al\mbox{.}(2018)]%
        {park2018adam}
\bibfield{author}{\bibinfo{person}{Seonwook Park}, \bibinfo{person}{Christoph Gebhardt}, \bibinfo{person}{Roman R{\"a}dle}, \bibinfo{person}{Anna~Maria Feit}, \bibinfo{person}{Hana Vrzakova}, \bibinfo{person}{Niraj~Ramesh Dayama}, \bibinfo{person}{Hui-Shyong Yeo}, \bibinfo{person}{Clemens~N Klokmose}, \bibinfo{person}{Aaron Quigley}, \bibinfo{person}{Antti Oulasvirta}, {et~al\mbox{.}}} \bibinfo{year}{2018}\natexlab{}.
\newblock \showarticletitle{Adam: Adapting multi-user interfaces for collaborative environments in real-time}. In \bibinfo{booktitle}{\emph{Proceedings of the 2018 CHI conference on human factors in computing systems}}. \bibinfo{pages}{1--14}.
\newblock


\bibitem[Schrepp et~al\mbox{.}(2023)]%
        {schrepp2023comparison}
\bibfield{author}{\bibinfo{person}{Martin Schrepp}, \bibinfo{person}{Jessica Kollmorgen}, {and} \bibinfo{person}{J{\"o}rg Thomaschewski}.} \bibinfo{year}{2023}\natexlab{}.
\newblock \showarticletitle{A Comparison of SUS, UMUX-LITE, and UEQ-S.}
\newblock \bibinfo{journal}{\emph{Journal of User Experience}} \bibinfo{volume}{18}, \bibinfo{number}{2} (\bibinfo{year}{2023}).
\newblock


\bibitem[Shah and Ghahramani(2016)]%
        {shah2016pareto}
\bibfield{author}{\bibinfo{person}{Amar Shah} {and} \bibinfo{person}{Zoubin Ghahramani}.} \bibinfo{year}{2016}\natexlab{}.
\newblock \showarticletitle{Pareto frontier learning with expensive correlated objectives}. In \bibinfo{booktitle}{\emph{International conference on machine learning}}. PMLR, \bibinfo{pages}{1919--1927}.
\newblock


\bibitem[Todi et~al\mbox{.}(2016)]%
        {todi2016sketchplore}
\bibfield{author}{\bibinfo{person}{Kashyap Todi}, \bibinfo{person}{Daryl Weir}, {and} \bibinfo{person}{Antti Oulasvirta}.} \bibinfo{year}{2016}\natexlab{}.
\newblock \showarticletitle{Sketchplore: Sketch and explore with a layout optimiser}. In \bibinfo{booktitle}{\emph{Proceedings of the 2016 ACM Conference on Designing Interactive Systems}}. \bibinfo{pages}{543--555}.
\newblock


\bibitem[Yamamoto et~al\mbox{.}(2022)]%
        {yamamoto2022photographic}
\bibfield{author}{\bibinfo{person}{Kenta Yamamoto}, \bibinfo{person}{Yuki Koyama}, {and} \bibinfo{person}{Yoichi Ochiai}.} \bibinfo{year}{2022}\natexlab{}.
\newblock \showarticletitle{Photographic lighting design with photographer-in-the-loop bayesian optimization}. In \bibinfo{booktitle}{\emph{Proceedings of the 35th annual ACM symposium on user interface software and technology}}. \bibinfo{pages}{1--11}.
\newblock


\end{thebibliography}

\appendix
\balance

\section{Task explanations}
\label{app:task-explanations}
We introduced participants of our study to the two tasks, free personalization and targeted personalization.
In addition, we gave them instructions on how to adjust UI elements manually. 
Below, we provide the instructions used in our study:

\begin{itemize}
    \item[] \textbf{Free personalization}:
    \\
    You are free to adjust UI widgets' position until you are satisfied with the overall layout.
    \item[] \textbf{Targeted personalization}:
    \\
    When adjusting the position of virtual UI elements and evaluating the updated layout, you should focus only on the following factors:

    \begin{itemize}
        \item \textbf{Shoulder Load} -- UI should be easily reachable without lifting your arms too high.
        \item \textbf{Neck Strain} -- UI should be at a comfortable eye level to avoid neck strain. (Neither too high nor too low)
        \item \textbf{Distance} -- UI should be within arm’s reach—neither too far nor too close.
        \item \textbf{Central Vision} -- UI should be positioned near the center of your field of view.
        \item \textbf{Semantic Agreement} -- UI elements and environmental objects can possess a semantic association, meaning that certain UI components are more semantically connected to specific environmental (physical) objects. For example, the UI of a video viewer is more semantically aligned with a laptop than with a book. This indicates that positioning the video viewer UI closer to the laptop rather than the book enhances semantic agreement.

    \end{itemize}
    \item[] \textbf{Manual UI element adjustments}:
    \\
    \cmanual{}: Review all UI widgets and adjust any that you're not satisfied with until you are satisfied with the layout.
    \\
    \cauit{}, \cours{}: Adjust 1-3 UI elements as an example. The tool will automatically refine the layout accordingly. If you're satisfied with the updated layout, you can stop; otherwise, continue refining elements.

\end{itemize}

\section{Scenario explanations}
\label{app:scenario-explanations}
We introduced participants of our study to the two scenarios, coffee shop and home office, with the following text:
\\
You will complete two tasks that involve customizing the positions of virtual UI elements. These tasks take place in two different scenarios:

\begin{itemize}
    \item[] \textbf{Relaxing in a coffee shop} -- Imagine you're seated at a table in a coffee shop. UI elements can be arranged in the space directly in front of you and to your left side. You arrange UI widgets for an enjoyable browsing experience using apps like Instagram, a video viewer, a messenger, news websites, and a music player.
    Your goal is to adjust the position of these widgets to create the most comfortable and convenient UI layout for an enjoyable browsing experience.
    \item[] \textbf{Working from home} -- Imagine you're working from home, sitting at a desktop. UI elements can be positioned in the area in front of you, as well as to your left and right sides. You’ll organize UI elements for efficient literature research using research papers, search engines, Slack, a messaging app, and a music player.
    Your goal is to arrange these widgets to create the most comfortable and convenient UI layout for your literature search.
\end{itemize}

\end{document}